\documentclass[reprint,superscriptaddress,amsmath,amssymb,aps]{revtex4-2}

\usepackage[version=3]{mhchem} 
\usepackage{hyperref}
\usepackage{amsfonts}
\usepackage{amsmath}
\usepackage{amssymb}
\usepackage{graphicx}
\usepackage{xcolor}

\usepackage{tabulary}
\usepackage{colortbl}

\usepackage[utf8]{inputenc}
\usepackage[T1]{fontenc}

\usepackage{color}
\usepackage{listings}
\usepackage{xpatch}

\usepackage{array}
\newcommand\Tstrut{\rule{0pt}{2.6ex}}         
\newcommand\Bstrut{\rule[-0.9ex]{0pt}{0pt}}   

\begin{document}

\title{Stochastic magnetic actuated random transducer devices based on perpendicular magnetic tunnel junctions}

\author{L. Rehm}
\email{laura.rehm@nyu.edu}
\affiliation{Center for Quantum Phenomena, Department of Physics, New York University, New York, New York 10003, USA}
\author{C. C. M. Capriata}
\affiliation{Center for Quantum Phenomena, Department of Physics, New York University, New York, New York 10003, USA}
\affiliation{Division of Electronics and Embedded Systems, KTH Royal Institute of Technology, 10044 Stockholm, Sweden}
\author{S. Misra}
\affiliation{Sandia National Laboratories, Albuquerque, New Mexico 87185, USA}
\author{J. D. Smith}
\affiliation{Sandia National Laboratories, Albuquerque, New Mexico 87185, USA}
\author{M. Pinarbasi}
\affiliation{Spin Memory Inc., Fremont, California 94538, USA}
\author{B. G. Malm}
\affiliation{Division of Electronics and Embedded Systems, KTH Royal Institute of Technology, 10044 Stockholm, Sweden}
\author{A. D. Kent}
\email{andy.kent@nyu.edu}
\affiliation{Center for Quantum Phenomena, Department of Physics, New York University, New York, New York 10003, USA}

\date{\today}

\begin{abstract} 
True random number generators are of great interest in many computing applications such as cryptography, neuromorphic systems and Monte Carlo simulations. Here we investigate perpendicular magnetic tunnel junction nanopillars (pMTJs) activated by short duration (ns) pulses in the ballistic limit for such applications. In this limit, a pulse can transform the Boltzmann distribution of initial free layer magnetization states into randomly magnetized down or up states, i.e. a bit that is 0 or 1, easily determined by measurement of the junction's tunnel resistance. It is demonstrated that bitstreams with millions of events: 1) are very well described by the binomial distribution; 2) pass multiple statistical tests for true randomness, including all the National Institute of Standards tests for random number generators with only one XOR operation; 3) can be used to create a uniform distribution of 8-bit random numbers;  and 4) can have no drift in the bit probability with time. The results presented here show that pMTJs operated in the ballistic regime can generate true random numbers at GHz bitrates, while being more robust to environmental changes, such as their operating temperature, compared to other stochastic nanomagnetic devices.
\end{abstract}

\maketitle

\section{\label{sec:level1}Introduction}
True random number generators (TRNGs) are of great interest for many applications such as cryptography~\cite{McInnes1991}, neuromorphic systems~\cite{Schuman2017}, and Monte Carlo simulations~\cite{Harrison2010}, which are extensively used to model and solve complex problems like accurate climate models, biological processes and particle production in high-energy colliders.
There are TRNGs based on microscopic phenomena, such as thermal noise~\cite{Huang2001}, quantum fluctuations ~\cite{Herrero_Collantes2017} ---  such as radioactive decay~\cite{rohe2003randy} --- and atmospheric environmental noise~\cite{Kumar2010}. Nonetheless, an important goal remains the discovery and development of TRNGs devices that are equally compact, fast, energy efficient, and robust with respect to device-to-device variability and environmental changes, such as their operating temperature.

Magnetic noise represents a new opportunity in this regard as small ferromagnetic elements can be two-state systems, with their magnetization ``up'' and ``down'' states separated by an energy barrier $E_b$. If the energy barrier is comparable to the thermal energy $kT$, where $k$ is the Boltzmann constant and $T$ is the device operating temperature, the magnetization fluctuates between the two states, which is known as superparamagnetism. A magnetic tunnel junction (MTJ) device can convert these thermally driven magnetization fluctuations into two-level electrical signals that are easily read out and, further, MTJs are readily integrated with complementary metal-oxide semiconductor (CMOS) technology see e.g.~\cite{Prenat2007, Matsunaga2008, Zhao2008, Kent2015, Deng2016, Kumar2019, Barla2020ALU, Barla2020}. In fact, MTJs with a superparamagnetic magnetic layer have already been explored for probabilistic computing~\cite{Vodenicarevic2017, Borders2019, Kaiser2019,Hayakawa2021,Safranski2021,Parks2018}. However, these devices are either passive or driven by a constant bias and as a result their response depends strongly on the environmental noise. Their rate of fluctuations can be described by the N\'{e}el-Brown formula $\Gamma = \Gamma_0 \exp(-E_b/kT)$, where $\Gamma_0$ is the attempt frequency~\cite{Neel1949,Brown1963}. This indicates that the fluctuation rate is extremely (exponentially) sensitive to the temperature as well as changes in $E_b$ associated with device and material parameter variations and external influences, such as magnetic fields.

Another phenomenon shown to be of interest for random number generation is the stochasticity of spin-transfer-torque (STT) switching of MTJs~\cite{SLONCZEWSKI1996L1,Berger1996,BERTOTTI2009271,Albert2002}. STT devices are conventionally used in memory applications, for which they are engineered to have two stable magnetic states~\cite{Kent2015,Slaughter2012,Thomas2015} --- high energy barriers ($E_b>60kT$) --- but also show great promise for this application due to their small device foot print ($\leq 20$~nm)~\cite{Jinnai2021}, energy-efficiency (fJ), fast operation (sub-ns), and controllability through their voltage bias or pulse time~\cite{Rehm2019}. While they have been investigated in the thermally assisted spin transfer regime (long pulse limit)~\cite{Yuasa2013,Fukushima2014} and with pulse durations approaching the thermally assisted spin transfer regime~\cite{Choi2014}, their operation in the ballistic switching limit (low-ns duration pulses) has yet to be explored. 
In the ballistic limit the resulting junction state and thus the resulting bit (0 or 1) is random mainly because of the Boltzmann distribution of initial magnetization states. 

\begin{figure*}
\includegraphics[width=0.98\textwidth,keepaspectratio]{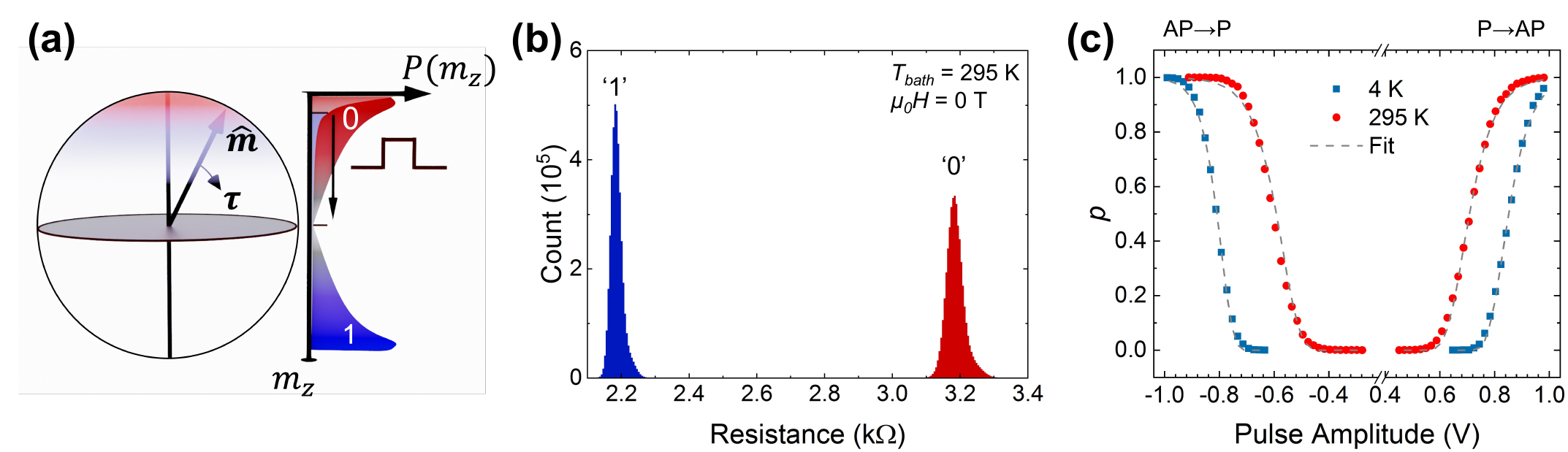}
\caption{Spin-transfer-torque switching in the ballistic regime. {\bf (a)} Schematic of the Bloch sphere with an initial magnetization represented by a vector labeled $\hat{m}$ (left) and the initial thermal magnetization distribution indicated in light red and blue colors close to the north pole. After a current pulse the magnetization's $z$-projection is bimodal, indicated schematically in dark red (0) and blue (1) on the right. {\bf (b)}  Histogram of the measured resistance values of a 40~nm diameter pMTJ for 8M switching attempts with switching probability $p \approx 0.5$ for the AP$\rightarrow$P transition at $T_\mathrm{bath}$ = 295~K and zero external field. The left peak corresponds to the resistance values where the device switched into P state (bit 1) and the right peak corresponds to the resistance values where the device remained in its initial AP state (bit 0). {\bf (c)} Switching probability versus pulse amplitude for a 1-ns-long pulse at 4~K (blue squares) and 295~K (red dots) with zero applied field. Each point is an average of 10,000 switching trials. The dashed gray curves are fits to the model described in the main text.}
\label{Fig:1}
\end{figure*}

We denote this a stochastic magnetic actuated random transducer (SMART) device because the pulse activates the junction to generate a random bitstream, much like a coin flip. Figure~\ref{Fig:1}(a) shows the Bloch sphere with an initial magnetization represented by a vector labeled $\hat{m}$. The blue/red shaded area represents the thermal distribution of the initial magnetization states. The corresponding Boltzmann distribution of the initial magnetization z-component ($m_z$) is represented on the right. After applying a pulse, blue regions will relax to $m_z = -1$  (switched, a bit 1) and red to $m_z = +1$ (not switched, a bit 0), the initial magnetization state.

Here we analyze and experimentally demonstrate the probabilistic behavior of medium energy barrier ($E_b \simeq 39 kT$) perpendicularly magnetized MTJs (pMTJs) in the ballistic limit. We will show that the stochastic nature of the STT switching of these devices operating at a switching probability of 50\% can be very well described by the statistics of Bernoulli trials. By whitening the experimentally obtained data stream with only one XOR operation~\cite{Matsui} we fully pass the National Institute of Standards and Technology (NIST) statistical test suite for random number generators~\cite{NIST}. In addition, we use the same bitstream to generate a uniform distribution of random numbers, which is important for many applications.

\section{Switching Probability in the Ballistic Limit}
Ballistic switching refers to the spin-torque-induced magnetization dynamics for short current or voltage pulses, typically on the order or less than a few nanoseconds. The main idea is that the pulse transfers angular momentum to the magnetic layer and there is no (or little) time for magnetic fluctuations during the pulse. The pulse can be considered to amplify the Boltzmann distribution of initial magnetization states, as shown schematically in Fig.~\ref{Fig:1}(a). The switching probability in a macrospin model in the ballistic limit is given by~\cite{Butler2012, Munira2012, Liu2014}:
\begin{equation}
p  = \exp \left\{-\frac{\pi^2 \Delta_\mathrm{eff}}{4} \exp\left[-\left(\frac{V}{V_{c0}} -
1\right)\frac{2\tau}{\tau_D} \right] \right\}, 
\label{Eq:Pswitch} 
\end{equation}
where $\Delta_\mathrm{eff}$ is the effective stability factor, which depends inversely on temperature and converges to $\Delta=E_b/(kT)$ in the very short duration-high amplitude pulse limit. $V_{c0}$ is the threshold bias, the switching threshold in the long pulse limit and $\tau$ is the applied pulse duration. $\tau_D$ is the intrinsic time scale for the dynamics $\tau_D = (1 + \alpha^ 2)/(\alpha \gamma \mu_0 H_k)$ ~\cite{Liu2014}, with $\alpha$ the damping constant, $\gamma$ the gryomagnetic ratio, and $\mu_0$ the permeability of free space. $H_k$ is the anisotropy field: $H_k \equiv 2K_p/\mu_0 M_s -M_s$. Here $K_p$ is the perpendicular anisotropy and $M_s$ is the saturation magnetization. Hence, $\tau_D$ is only indirectly dependent on temperature through changes in material parameters with temperature. The same is the case for $V_{c0}$, it depends on material parameters that are a function of temperature~\cite{Rehm2021}.   

As we aim to operate the device near a switching probability $p$ of $0.5$, we derive an expression for the switching probability close to this value.
With a linear approximation of Eq.~\ref{Eq:Pswitch} around $p$= 0.5 we find 
\begin{eqnarray}
    p(V)&=&\frac{1}{2}+\frac{\tau \ln 2}{\tau_D V_{c0}}(V-V_{1/2}),
    \label{Eq:P12} \\
 V_{1/2}&=&V_{c0}+\frac{\tau_D V_{c0}}{2 \tau}\ln \left( \frac{\pi^2 \Delta_\mathrm{eff}}{4 \ln2}\right),
\label{Eq:V0p5}
\end{eqnarray}
where $V_{1/2}$ is the 50\% switching voltage threshold.
Therefore, the switching voltage for $p$ = 0.5 only depends logarithmically on temperature through $\Delta_\mathrm{eff}$ and on material parameter $V_{c0}$ and $\tau_D$ that do not vary greatly with temperature (for device operation well below the magnet's Curie temperature).

\section{SMART Device Characteristics}
We conducted experiments in the ballistic limit on circularly shaped pMTJs with a MgO tunnel barrier. The free layer is a composite CoFeB/W/CoFeB layer stack (with composition CoFeB $\equiv$ \ce{Co18Fe54B28}) and the pinned layer is a CoFeB layer that is ferromagnetically coupled to a synthetic antiferromagnet layer structure composed of two Pt/Co multilayers separated by a thin Ru layer. The room-temperature $\Delta$ of the pMTJs studied is 39 and the resistance-area product is $\simeq 3~\Omega \mu$m\textsuperscript{2} with a free layer diameter of 40~nm. A more detailed description of the devices can be found in Ref.~\cite{Rehm2019,Rehm2021}. 

The stochastic write behavior of our SMART devices was studied by repeatedly applying a reset-read-write-read scheme. We first reset to the desired state with a 50~$\mu$s-long pulse and pulse amplitudes well above the switching voltage (see Fig.~1(c) in the Supplemental Material~\cite{SMCF2022}) and then we determine the state of the device before and after a 1~ns write pulse with a data acquisition (DAQ) board (National Instruments PCIe-6353). We chose a write pulse duration of 1~ns (Tektronix AWG 7102) as this is close to the most energy efficient device switching condition, corresponding to pulse times $\simeq \tau_D$~\cite{Rehm2019}.

Figure~\ref{Fig:1}(b) shows a histogram of the measured resistance values for 8 million switching attempts obtained at room temperature ($T_\mathrm{bath}$ = 295~K) and zero external field. We can clearly observe two distinct, well-separated resistance distributions corresponding to parallel (P) and antiparallel (AP) magnetization configurations with $\langle R_{P}\rangle$ and $\langle R_{AP}\rangle$ of 2,180 and 3,169~$\Omega$, respectively. Based on the resistance distribution we can then further evaluate if the resistance value obtained during the second read pulse falls into the criteria $\langle R_{AP}\rangle  \pm 5\times\sigma_{AP}$  or $ \langle R_{P} \rangle  \pm 5\times\sigma_{P}$, where $\langle R \rangle$ is the mean of a resistance distribution and $\sigma$ is its standard deviation. Thus, we assess if the device either switched or did not switch, assigning a 1 or 0 respectively. Additional device characterization can be found in Sec.~1 of the Supplemental Material~\cite{SMCF2022}.

\section{Results and Analysis}
Figure~\ref{Fig:1}(c) shows the switching probability as a function of pulse amplitude at room temperature (red dots). Each point in the graph represents $N_T = 10,000$ switching attempts. As expected, the switching probability increases monotonically with pulse amplitude. This data is analyzed with the ballistic macrospin model to obtain key device parameters. The variation of the switching probability with pulse amplitude at $p=1/2$, i.e. $dp/dV|_{p=1/2}$ gives $\tau_D V_{c0}$ (Eq.~\ref{Eq:P12}) and the pulse amplitude to achieve $p=0.5$ is $V_{1/2}$. With $V_{1/2}$ and $\tau_D V_{c0}$ determined, Eq.~\ref{Eq:V0p5} provides a relation between $V_{c0}$ and $\Delta_\mathrm{eff}$. We thus fit our $p(V)$ data with the expression for the switching probability (Eq.~\ref{Eq:Pswitch}) with only one fit parameter, $\Delta_\mathrm{eff}$.
The fits can be seen in Fig.~\ref{Fig:1}(c) as the gray dashed lines. We find that they capture the characteristics of the experimental data well. In addition, we repeated the same measurements and analysis at low temperature, $T_\mathrm{bath}$ = 4~K (blue squares, Fig.~\ref{Fig:1}(c)), to investigate the effect of a large change in temperature on the switching probability and fit parameters~\footnote{It should be noted during pulses the temperature of the device can be significantly larger than the bath temperature due to junction heating~\cite{Rehm2021}.}. The fit parameters are given in Table~\ref{t:Fit}. Interestingly, $\Delta_\mathrm{eff}$ does not change significantly with temperature, nor does $\tau_D V_{c0}$, while $V_{c0}$ depends more strongly on temperature. We used the junction material parameters to compute $\tau_D V_{c0}$ in the macrospin model and find that it is generally a factor of 3 to 8 larger than what is found experimentally. We further find $\Delta_\mathrm{eff}$ to be much smaller than $\Delta$ determined from measurements in the long-pulse limit~\cite{Rehm2021}. We discuss these characteristics below.

\begin{table*}
\caption{Fit parameters $\Delta_\mathrm{eff}$ and $V_{c0}$ with their corresponding standard deviations as well as $\tau_D V_{c0}$ obtained by fitting the data and from a macrospin model.}
\label{t:Fit}
\setlength{\tabcolsep}{10pt}
\begin{tabular}{lccccc}
\noalign{\smallskip} \hline \hline \noalign{\smallskip}
T & Transition & $\Delta_\mathrm{eff}$ & $V_{c0}$ & Fit: $\tau_D V_{c0}$ & Macrospin: $\tau_D V_{c0}$\\
(K) &  &  & (V) & (sV) & (sV)\Bstrut\\
\hline
4 & AP$\rightarrow$P & 1.600 $\pm$ 2$\times10^{-3}$ & -0.8542 $\pm$ 4$\times10^{-5}$ & 0.87$\times10^{-10}$ & 6.9$\times10^{-10}$\Tstrut\\
4 & P$\rightarrow$AP & 1.585 $\pm$ 4$\times10^{-4}$ & 0.9094 $\pm$ 1$\times10^{-5}$ & 1.11$\times10^{-10}$ & 6.9$\times10^{-10}$\\
295 & AP$\rightarrow$P & 1.475 $\pm$ 5$\times10^{-4}$ & -0.6626 $\pm$ 2$\times10^{-5}$ & 1.27$\times10^{-10}$ & 5.66$\times10^{-10}$\\
295 & P$\rightarrow$AP & 1.585 $\pm$ 8$\times10^{-4}$ & 0.7732 $\pm$ 3$\times10^{-5}$ & 1.25$\times10^{-10}$ & 5.66$\times10^{-10}$\Bstrut\\
\noalign{\smallskip} \hline \hline \noalign{\smallskip}
\end{tabular}
\end{table*}

\begin{figure*}
\includegraphics[width=0.98\textwidth,keepaspectratio]{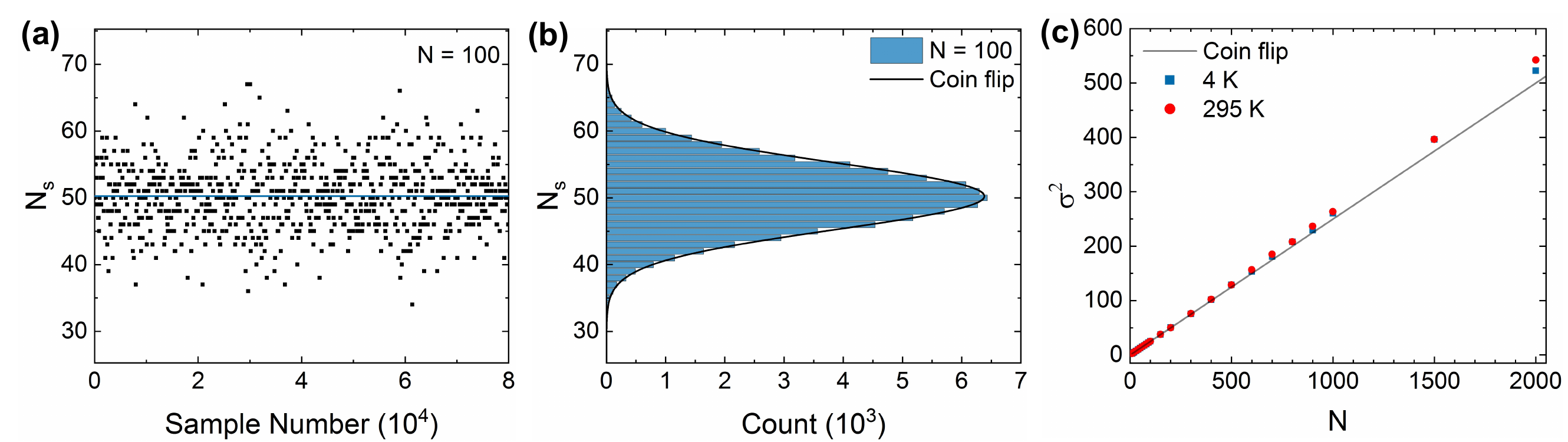}
\caption{Statistical analysis of the number of switched attempts. {\bf (a)} Number of switched attempts $N_s$ for the AP$\rightarrow$P transition in a sample size of N = 100 versus the sample number for the same device at $T_\mathrm{bath}$ = 295~K. The solid straight line in the plot represents the switching probability $p = 0.5027$ of the whole data set ($N_T \approx 8\times10^6$). {\bf (b)} Histogram of the switched events $N_s$ of the same data set. The solid black line in the plot shows the binomial distribution of a slightly weighted coin flip with a probability of $p$ = 0.5027 and a variance of $\sigma^2=$ 25.00. {\bf (c)} Variance $\sigma^2$ of the switched attempts distribution versus the number of trials in a sample $N$ at $T_\mathrm{bath}$ = 4~K (blue squares) and 295~K (red dots). The solid black line graph shows the expectation for a Bernoulli process with $p=0.5$.} 
\label{Fig:2}
\end{figure*}

To further investigate the probabilistic behavior of our pMTJs, we increased the number of switching attempts $N_T$ to 8 million and focused on $p \simeq 0.5$. To analyze the switching statistics versus time and generate a switching probability distribution we consider non-overlapping $N=100$ trials as samples. We then count the successfully switched attempts $N_s$ in each sample (based on the criteria described earlier). The results for the AP$\rightarrow$P transition at room temperature are shown in Fig.~\ref{Fig:2}(a). The solid straight blue line represents the data set's probability of $p = 0.5027$. We observe no drift. Figure~\ref{Fig:2}(b) shows a number of switching events histogram; we observe a distribution that is symmetric around the mean. 

Bernoulli trials describe a random event with two possible outcomes. Each trial is independent and the probability of outcomes does not change over trials.
If the outcomes are `heads' and `tails,' and the probability of `heads' is $p=0.5$, then the Bernoulli trials describe flips of a fair coin. Letting $p$ represent the probability of `success' and $N$ represent the number of trials, for large $N$ the number of successes $x$ is approximately normally distributed. The probability density function (PDF) for the number of successes is given by:
\begin{equation}
  f(x) =\frac{1}{\sqrt{2\pi\sigma^2}}\exp\Big\{-\frac{(x-\mu)^2}{2\sigma^2} \Big\},
  \label{eqn:1}
\end{equation}
where $\mu= pN$ is the mean and $\sigma= \sqrt{N(1-p)}$ is the standard deviation. Returning to the pMTJs, of  $N_T = 7,999,401$ total attempts, we found $50.27$\% were bit flips. Using the Bernoulli trial framework, our outcomes are whether or not a bit flip occurred and we take $p=0.5027$. Then, for $N = 100$ trials the expected mean number of bit flips is $\mu = 50.27$ with variance $\sigma^2= 25.00$. The solid black line in Fig.~\ref{Fig:2}(b) is a plot of the normal distribution (Eq.~\ref{eqn:1}) with these values; the histogram of our data is clearly very well characterized by this distribution function.

We continued by calculating the mean of the switched attempts $\mu$ = $ \langle N_s \rangle$ as well as the variance of the corresponding distribution as a function of $N$, the number of trials in a sample, with $N$ varying from $10$ to $2000$:  
\begin{equation}
  \sigma^2 = \langle N_s^2 \rangle -  \langle N_s \rangle^2.
  \label{eqn:4}
\end{equation}
This is shown in Fig.~\ref{Fig:2}(c) as the red dots for the room temperature data and the blue squares for data at 4~K. The solid black line represents the expected variance as a function of $N$ for $p=0.5$. The data deviates upward from the line at large sample size $N$, as expected when the number of total samples is not sufficiently large (i.e. when $N$ approaches $N_T$).

\section{Sampling a Uniform Distribution}
\begin{figure*}
\includegraphics[width=0.98\textwidth,keepaspectratio]{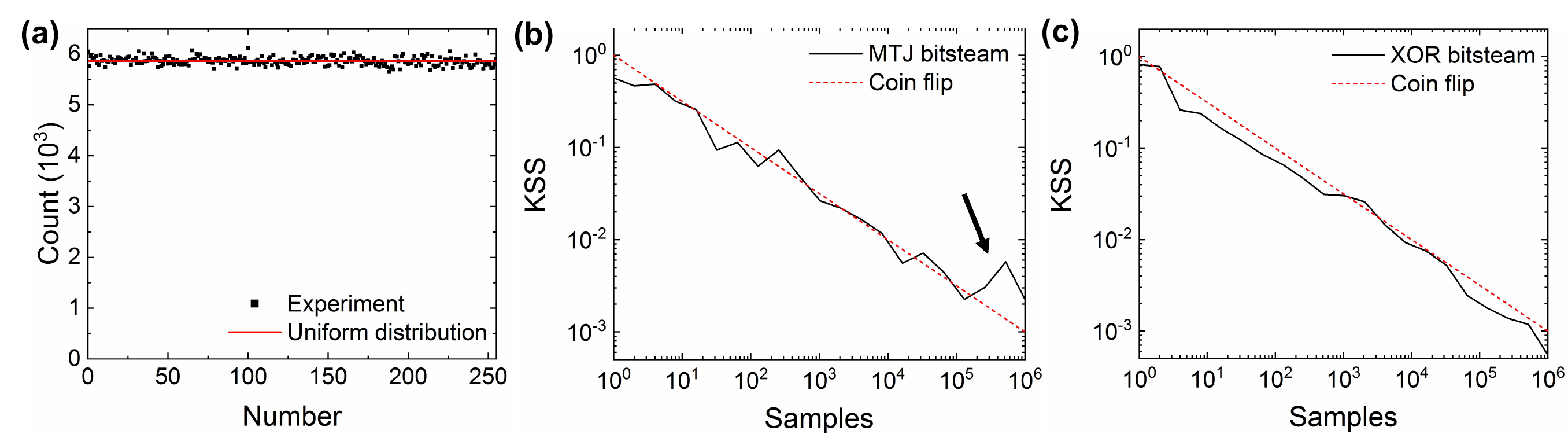}
\caption{Uniform distribution. {\bf (a)} 256-channel histogram of 1,499,996 8-bit values generated using the SMART bitstream.  The histogram was fit to a constant and provides a $\chi^2$/DOF = 1.13. {\bf (b)} KSS versus the number of uniform samples for a bitstream generated using the pure  bitstream (solid black line) and {\bf (c)} the bitstream whitened by taking the XOR of consecutive bits (solid black line) with the dashed red line representing the result of an ideal distribution (coin flip) with KSS = 1/$\sqrt{N}$. The black arrow in (b) shows where deviations from the ideal distribution occur due to small probability deviations from $p=0.5$.} 
\label{Fig:3}
\end{figure*}

A critical elementary operation in probabilistic computing applications, e.g. Monte Carlo simulations, is to draw samples from different distributions. The fitness of different devices can be evaluated in terms of the statistical quality of the samples their bitstreams generate. We place eight consecutive readings of our SMART device in each position of an 8-bit string, interpreting it as a random number from 0 through 255. Figure~\ref{Fig:3}(a) shows the histogram for 1,499,996 8-bit samples created from $N_T = 11,999,968$ measurements of the P$\rightarrow$AP transition at $T_\mathrm{bath}$ = 295~K. The experimentally obtained random numbers (black data points) can be very well described by an ideal uniform distribution (red solid line) and provide a $\chi^2$/DOF = 1.13, where DOF stands for the degrees of freedom defined as the number of data points minus 1. 

Transformation of samples from a uniform distribution to various standard distributions, and repeated calculations based on drawing many samples, both rely on the quality of the uniform random samples generated. Using a device that produces independent and identically distributed coin flips with $p = 0.5$ to generate an infinite number of uniform samples should yield a PDF that is a constant, and a cumulative distribution function (CDF) that is a line with a constant slope. We quantify the error from a finite set of $N$ samples using the Kolmogorov-Smirnov statistic (KSS), which measures the maximum deviation of the empirical CDF of the observed distribution from the desired distribution ~\cite{thas2010comparing}. For a set of $N$ uniform samples, the KSS $\sim$ 1/$\sqrt{N}$ (Fig.~\ref{Fig:3}(b), dashed red line). We now interpret the 8-bit string as a uniformly distributed random number [0,1) in 8-bit fixed point notation. A plot of the KSS as a function of $N$ uniform samples generated using the SMART bitstream, shown in Fig.~\ref{Fig:3}(b), shows a deviation from 1/$\sqrt{N}$ behavior for $N \sim 10^5$. This is a consequence of $p =0.5+\delta$ with $\delta$ = 2$\times10^{-3}$ for this data set. Upon whitening the data by taking the logic XOR, exclusive or function, of consecutive bits~\cite{Matsui}, the error away from $p=0.5$ is reduced to 4$\times10^{-4}$, and the KSS follows $1/\sqrt{N}$ to $N > 10^5$ (Fig.~\ref{Fig:3}c). More generally, any deviation from the desired $p$ by $\delta$ will produce a lower limit beyond which increased sampling will not improve the KSS beyond order $\delta$. Heuristically, there is a limit to the number of samples we can take while ignoring the small deviation from a fair flip. This derives from the deviation in the probability of most 8-bit strings being of order $\delta$, with the deviation being positive if the number of 1s is more than the number of 0s, vice versa, and zero if they are equal. Put differently, the number of samples beyond which deviation of the generated samples from the desired distribution becomes statistically significant is of order 1/$\delta^2$. 

We close this discussion by emphasizing that there are other characteristics of the SMART bitsteam, such as an absence of correlations, and timing that indicates random draws, which lead to the relationship of $p$ and the KSS. In devices where those pathologies are present, even if they more faithfully produce $p=0.5$, limitations to the quality of the samples will simply have a different origin. Analysis of these connections is provided in greater detail in Sec.~2 in the Supplemental Material~\cite{SMCF2022}. 

\section{NIST Statistical Tests}
In addition to the statistical analysis of the STT switching of our pMTJs and the generation of uniform random numbers, we also tested the generated data streams with the NIST statistical test suite~\cite{NIST}. The test suite consists of numerous frequency and non-frequency related tests. The frequency related tests evaluate the number of 1s in the data set, while the non-frequency related tests check for special patterns for example. The results of a $N_T= 12$M data stream can be found in Table~\ref{t:NIST}.

\begin{table*}
\caption{NIST tests results of the SMART bitstreams. The whitened bitstreams after one and two XOR operations.}
\footnotesize
\label{t:NIST}
\begin{tabular}{ |p{6cm}||p{1.15cm}|p{1.15cm}|p{1.15cm}|p{1.15cm}|p{1.15cm}|p{1.15cm}| }
\hline
& \multicolumn{3}{c}{AP$\rightarrow$P} & \multicolumn{3}{|c|}{P$\rightarrow$AP}\\
\hline
\hline
Test name / XOR stages & 0 & 1 & 2 & 0 & 1 & 2\\
\hline
Frequency (Monobit) & \cellcolor{red}0/1 & \cellcolor{green}1/1 & \cellcolor{green}1/1 & \cellcolor{red}0/1 & \cellcolor{green}1/1 & \cellcolor{green}1/1\\
Frequency within a Block & \cellcolor{green}1/1 & \cellcolor{green}1/1 & \cellcolor{green}1/1 & \cellcolor{green}1/1 & \cellcolor{green}1/1 & \cellcolor{green}1/1\\
Run & \cellcolor{red}0/1 & \cellcolor{green}1/1 & \cellcolor{green}1/1 & \cellcolor{green}1/1 & \cellcolor{green}1/1 & \cellcolor{green}1/1\\
Longest Run of Ones in a Block & \cellcolor{green}1/1 & \cellcolor{green}1/1 & \cellcolor{green}1/1 & \cellcolor{green}1/1 & \cellcolor{green}1/1 & \cellcolor{green}1/1\\
Binary Matrix Rank & \cellcolor{green}1/1 & \cellcolor{green}1/1 & \cellcolor{green}1/1 & \cellcolor{green}1/1 & \cellcolor{green}1/1 & \cellcolor{green}1/1\\
Discrete Fourier Transform (Spectral) & \cellcolor{green}1/1 & \cellcolor{green}1/1 & \cellcolor{green}1/1 & \cellcolor{green}1/1 & \cellcolor{green}1/1 & \cellcolor{green}1/1\\
Non-Overlapping Template Matching & \cellcolor{green}148/148 & \cellcolor{green}148/148 & \cellcolor{green}148/148 & \cellcolor{yellow}147/148 & \cellcolor{green}148/148 & \cellcolor{green}148/148\\
Overlapping Template Matching & \cellcolor{green}1/1 & \cellcolor{green}1/1 & \cellcolor{green}1/1 & \cellcolor{green}1/1 & \cellcolor{green}1/1 & \cellcolor{green}1/1\\
Maurer's Universal Statistical  & \cellcolor{green}1/1 & \cellcolor{green}1/1 & \cellcolor{green}1/1 & \cellcolor{green}1/1 & \cellcolor{green}1/1 & \cellcolor{green}1/1\\
Linear Complexity & \cellcolor{green}1/1 & \cellcolor{green}1/1 & \cellcolor{green}1/1 & \cellcolor{green}1/1 & \cellcolor{green}1/1 & \cellcolor{green}1/1\\
Serial & \cellcolor{green}2/2 & \cellcolor{green}2/2 & \cellcolor{green}2/2 & \cellcolor{green}2/2 & \cellcolor{green}2/2 & \cellcolor{green}2/2\\
Approximate Entropy & \cellcolor{green}1/1 & \cellcolor{green}1/1 & \cellcolor{green}1/1 & \cellcolor{green}1/1 & \cellcolor{green}1/1 & \cellcolor{green}1/1\\
Cumulative Sums (Forward) & \cellcolor{red}0/1 & \cellcolor{green}1/1 & \cellcolor{green}1/1 & \cellcolor{red}0/1 & \cellcolor{green}1/1 & \cellcolor{green}1/1\\
Cumulative Sums (Reverse) & \cellcolor{red}0/1 & \cellcolor{green}1/1 & \cellcolor{green}1/1 & \cellcolor{red}0/1 & \cellcolor{green}1/1 & \cellcolor{green}1/1\\
Random Excursions & \cellcolor{green}8/8 & \cellcolor{green}8/8 & \cellcolor{green}8/8 & \cellcolor{green}8/8 & \cellcolor{green}8/8 & \cellcolor{green}8/8\\
Random Excursions Variant & \cellcolor{green}18/18 & \cellcolor{green}18/18 & \cellcolor{green}18/18 & \cellcolor{green}18/18 & \cellcolor{green}18/18 & \cellcolor{green}18/18\\
\hline
\end{tabular}
\end{table*}

We find that our initial data streams obtained at room temperature for the AP$\rightarrow$P transition with $p = 0.5027$ and P$\rightarrow$AP with $p = 0.5034$ pass 184 out of the 188 tests, resulting in an identical passing rate of 97.87\% (see XOR stage 0 in Table~\ref{t:NIST}). 
We whiten the bitstream with a single XOR operation and the effect of this process is again evident: the passing rate jumps to 100\%, all tests pass (see XOR stage 1 in Table~\ref{t:NIST}). While we are creating the bitstreams from a single device, it is clearly possible to parallelize the creation of the inputs by simply using multiple devices which will result in a reduction of the probability bias of the order $\delta$ to $\delta^2$, assuming identical device switching probabilities. See Sec.~3 in the Supplemental Material~\cite{SMCF2022} for more detailed information about the NIST tests.

\section{Discussion}
We have seen that the small probability bias of our SMART device can be handled by whitening the bitstream, but it can also be managed by simply adjusting the pulse amplitude or pulse duration. While the whitening of the bitstream is generally a great way to improve the quality of the random numbers, the latter option can only be used for a fixed (time-independent) deviation of the switching probability from 0.5. We expect our SMART devices to be relatively insensitive to changes in operating temperature. For instance, a temperature change of $\pm 15$~K would change $\Delta$ by $\pm 5\%$ and result in a change of $p \pm 3.5\%$ in the ballistic limit (Eq.~\ref{Eq:Pswitch}). However, for long pulses (i.e. $\tau \gg \tau_D$) the spin torque can be considered to modify the energy barrier to thermally activated reversal. In the paper by Fukushima {\em et al.}~\cite{Fukushima2014} they assume $p=1-\exp(- \Gamma_0 \tau \exp(-\Delta(1-v)^{\nu}))$, with $\nu=2$. Due to this double-exponential temperature dependence, the same change in $\Delta$ of $\pm 5\%$ would lead to $p \pm 47\%$ (assuming the same device parameters, $\Delta = 40$, $\Gamma_0=10^9$ s$^{-1}$, and $\tau=10^{-4}$~s). The same double-exponential temperature dependence can be found for the switching probability of superparamagnetic MTJs, where $p=1-\exp(- \Gamma_0 t \exp(-\Delta))$, and is thus very sensitive to small changes in temperature compared to the switching probability of our SMART devices.

We now return to device parameters found in fitting the switching probability versus pulse amplitude results in Fig.~\ref{Fig:1}(c), the results shown in Table~\ref{t:Fit}. First, we found $\Delta_\mathrm{eff} \ll \Delta$, with $\Delta$ determined by measurements with longer pulses~\cite{Rehm2021}. This is a characteristic of switching experiments with small overdrives, $V/V_{c0}$ greater than 1 but not much greater than 1. In this case thermal fluctuations during the pulse (not simply the initial magnetization state) play a role, as the spin torques amplify all fluctuations, even those during the pulse. This was found in a Fokker-Planck modeling of the switching process and in experiments on perpendicular spin-value nanopillars~\cite{Liu2014}. Importantly, in Ref.~\cite{Liu2014} it was shown that Eq.~\ref{Eq:Pswitch} still accurately characterizes the switching probability with a
$\Delta_\mathrm{eff}\ll \Delta$. As noted earlier, we also found that the fit values of $\tau_D V_{c0}$ are smaller than those expected based on the macrospin model with our device material parameters. This is consistent with other experiments on nanomagnetic switching dynamics that find shorter time scales $\tau_D$ than those in the macrospin model~\cite{Krause2009, Bedau2010}. Finally, the  change in the fit parameter $V_{c0}$ from $T_\mathrm{bath}=295$ to $4$~K is similar to the percentage changes we found in $V_{c0}$ in this temperature range by analyzing long-duration pulse results (see Fig.~3(a) of Ref.~\cite{Rehm2021}).

\section{Conclusion}
In summary, these results clearly demonstrate that medium energy barrier pMTJs operating in the ballistic limit have great potential for true random number generation. We have assessed the statistics of the experimentally obtained bitstreams by comparing it to a binomial distribution as well as explored their potential for numerous applications by sampling a uniform distribution. By whitening the bitstream with one XOR operation, we have shown that the bitstream passes all NIST tests designed for evaluating the quality of random number generators. In addition, compared to other nanomagnetic device concepts, such as STT devices operated with long duration pulses and superparamagnetic MTJs, we have found that our devices show the potential of having a lower temperature sensitivity, can have no drift of the switching probability with time and offer the possibility of precisely varying the switching probability by pulse conditions.

\begin{acknowledgements}
This paper describes objective technical results and analysis. Any subjective views or opinions that might be expressed in the paper do not necessarily represent the views of the U.S. Department of Energy or the United States Government. Sandia National Laboratories is a multimission laboratory managed and operated by National Technology \& Engineering Solutions of Sandia, LLC, a wholly owned subsidiary of Honeywell International Inc., for the U.S. Department of Energy’s National Nuclear Security Administration under contract DE-NA0003525.

We acknowledge support from the DOE Office of Science (ASCR/BES) Microelectronics Co-Design project COINFLIPS. We thank Shaloo Rakheja and Avik Ghosh for discussions of the ballistic macrospin model and Spin Memory for providing the magnetic tunnel junctions. We especially thank Bartek Kardasz for the layer stack deposition and characterization and Georg Wolf for the initial device screening. We would also like to thank Guanzhong Wu for his assistance in creating Fig.~1(a). This work was partially supported by the Swedish Research Council(VR), project Fundamental Fluctuations in Spintronics, 2017-04196. We also thank the funding agencies Nils and Hans Backmark Foundation (J-2021-2437) and Karl Engvers Foundation for supporting the project. This work was also partly funded under the Laboratory Directed Research and Development  program at Sandia National Laboratories.
\end{acknowledgements}

\bibliography{COINFLIP.bib}

\begin{thebibliography}{45}%
\makeatletter
\providecommand \@ifxundefined [1]{%
 \@ifx{#1\undefined}
}%
\providecommand \@ifnum [1]{%
 \ifnum #1\expandafter \@firstoftwo
 \else \expandafter \@secondoftwo
 \fi
}%
\providecommand \@ifx [1]{%
 \ifx #1\expandafter \@firstoftwo
 \else \expandafter \@secondoftwo
 \fi
}%
\providecommand \natexlab [1]{#1}%
\providecommand \enquote  [1]{``#1''}%
\providecommand \bibnamefont  [1]{#1}%
\providecommand \bibfnamefont [1]{#1}%
\providecommand \citenamefont [1]{#1}%
\providecommand \href@noop [0]{\@secondoftwo}%
\providecommand \href [0]{\begingroup \@sanitize@url \@href}%
\providecommand \@href[1]{\@@startlink{#1}\@@href}%
\providecommand \@@href[1]{\endgroup#1\@@endlink}%
\providecommand \@sanitize@url [0]{\catcode `\\12\catcode `\$12\catcode
  `\&12\catcode `\#12\catcode `\^12\catcode `\_12\catcode `\%12\relax}%
\providecommand \@@startlink[1]{}%
\providecommand \@@endlink[0]{}%
\providecommand \url  [0]{\begingroup\@sanitize@url \@url }%
\providecommand \@url [1]{\endgroup\@href {#1}{\urlprefix }}%
\providecommand \urlprefix  [0]{URL }%
\providecommand \Eprint [0]{\href }%
\providecommand \doibase [0]{https://doi.org/}%
\providecommand \selectlanguage [0]{\@gobble}%
\providecommand \bibinfo  [0]{\@secondoftwo}%
\providecommand \bibfield  [0]{\@secondoftwo}%
\providecommand \translation [1]{[#1]}%
\providecommand \BibitemOpen [0]{}%
\providecommand \bibitemStop [0]{}%
\providecommand \bibitemNoStop [0]{.\EOS\space}%
\providecommand \EOS [0]{\spacefactor3000\relax}%
\providecommand \BibitemShut  [1]{\csname bibitem#1\endcsname}%
\let\auto@bib@innerbib\@empty
\bibitem [{\citenamefont {McInnes}\ and\ \citenamefont
  {Pinkas}(1991)}]{McInnes1991}%
  \BibitemOpen
  \bibfield  {author} {\bibinfo {author} {\bibfnamefont {J.~L.}\ \bibnamefont
  {McInnes}}\ and\ \bibinfo {author} {\bibfnamefont {B.}~\bibnamefont
  {Pinkas}},\ }\bibfield  {title} {\bibinfo {title} {{On the Impossibility of
  Private Key Cryptography with Weakly Random Keys}},\ }in\ \href
  {https://doi.org/10.1007/3-540-38424-3_31} {\emph {\bibinfo {booktitle}
  {{Advances in Cryptology-CRYPTO' 90}}}},\ \bibinfo {editor} {edited by\
  \bibinfo {editor} {\bibfnamefont {A.~J.}\ \bibnamefont {Menezes}}\ and\
  \bibinfo {editor} {\bibfnamefont {S.~A.}\ \bibnamefont {Vanstone}}}\
  (\bibinfo  {publisher} {Springer Berlin Heidelberg},\ \bibinfo {address}
  {Berlin, Heidelberg},\ \bibinfo {year} {1991})\ pp.\ \bibinfo {pages}
  {421--435}\BibitemShut {NoStop}%
\bibitem [{\citenamefont {Schuman}\ \emph {et~al.}(2017)\citenamefont
  {Schuman}, \citenamefont {Potok}, \citenamefont {Patton}, \citenamefont
  {Birdwell}, \citenamefont {Dean}, \citenamefont {Rose},\ and\ \citenamefont
  {Plank}}]{Schuman2017}%
  \BibitemOpen
  \bibfield  {author} {\bibinfo {author} {\bibfnamefont {C.~D.}\ \bibnamefont
  {Schuman}}, \bibinfo {author} {\bibfnamefont {T.~E.}\ \bibnamefont {Potok}},
  \bibinfo {author} {\bibfnamefont {R.~M.}\ \bibnamefont {Patton}}, \bibinfo
  {author} {\bibfnamefont {J.~D.}\ \bibnamefont {Birdwell}}, \bibinfo {author}
  {\bibfnamefont {M.~E.}\ \bibnamefont {Dean}}, \bibinfo {author}
  {\bibfnamefont {G.~S.}\ \bibnamefont {Rose}},\ and\ \bibinfo {author}
  {\bibfnamefont {J.~S.}\ \bibnamefont {Plank}},\ }\bibfield  {title} {\bibinfo
  {title} {{A Survey of Neuromorphic Computing and Neural Networks in
  Hardware}},\ }\href {http://arxiv.org/abs/1705.06963} {\bibfield  {journal}
  {\bibinfo  {journal} {CoRR}\ }\textbf {\bibinfo {volume} {abs/1705.06963}}
  (\bibinfo {year} {2017})}\BibitemShut {NoStop}%
\bibitem [{\citenamefont {Harrison}(2010)}]{Harrison2010}%
  \BibitemOpen
  \bibfield  {author} {\bibinfo {author} {\bibfnamefont {R.~L.}\ \bibnamefont
  {Harrison}},\ }\bibfield  {title} {\bibinfo {title} {{Introduction to Monte
  Carlo Simulation}},\ }\href {https://doi.org/10.1063/1.3295638} {\bibfield
  {journal} {\bibinfo  {journal} {AIP Conference Proceedings}\ }\textbf
  {\bibinfo {volume} {1204}},\ \bibinfo {pages} {17} (\bibinfo {year}
  {2010})}\BibitemShut {NoStop}%
\bibitem [{\citenamefont {Zhun}\ and\ \citenamefont
  {Hongyi}(2001)}]{Huang2001}%
  \BibitemOpen
  \bibfield  {author} {\bibinfo {author} {\bibfnamefont {H.}~\bibnamefont
  {Zhun}}\ and\ \bibinfo {author} {\bibfnamefont {C.}~\bibnamefont {Hongyi}},\
  }\bibfield  {title} {\bibinfo {title} {A truly random number generator based
  on thermal noise},\ }in\ \href {https://doi.org/10.1109/ICASIC.2001.982700}
  {\emph {\bibinfo {booktitle} {ASICON 2001. 2001 4th International Conference
  on ASIC Proceedings (Cat. No.01TH8549)}}}\ (\bibinfo {year} {2001})\ pp.\
  \bibinfo {pages} {862--864}\BibitemShut {NoStop}%
\bibitem [{\citenamefont {Herrero-Collantes}\ and\ \citenamefont
  {Garcia-Escartin}(2017)}]{Herrero_Collantes2017}%
  \BibitemOpen
  \bibfield  {author} {\bibinfo {author} {\bibfnamefont {M.}~\bibnamefont
  {Herrero-Collantes}}\ and\ \bibinfo {author} {\bibfnamefont {J.~C.}\
  \bibnamefont {Garcia-Escartin}},\ }\bibfield  {title} {\bibinfo {title}
  {Quantum random number generators},\ }\href
  {https://doi.org/10.1103/RevModPhys.89.015004} {\bibfield  {journal}
  {\bibinfo  {journal} {Rev. Mod. Phys.}\ }\textbf {\bibinfo {volume} {89}},\
  \bibinfo {pages} {015004} (\bibinfo {year} {2017})}\BibitemShut {NoStop}%
\bibitem [{\citenamefont {Rohe}\ \emph {et~al.}(2003)\citenamefont {Rohe} \emph
  {et~al.}}]{rohe2003randy}%
  \BibitemOpen
  \bibfield  {author} {\bibinfo {author} {\bibfnamefont {M.}~\bibnamefont
  {Rohe}} \emph {et~al.},\ }in\ \href
  {https://citeseerx.ist.psu.edu/viewdoc/download?doi=10.1.1.110.9725&rep=rep1&type=pdf}
  {\emph {\bibinfo {booktitle} {{RANDy-A true-random generator based on
  radioactive decay}}}}\ (\bibinfo  {publisher} {Citeseer},\ \bibinfo {year}
  {2003})\ pp.\ \bibinfo {pages} {1--36}\BibitemShut {NoStop}%
\bibitem [{\citenamefont {Kumar}\ \emph {et~al.}(2020)\citenamefont {Kumar},
  \citenamefont {Jadhav}, \citenamefont {Misra},\ and\ \citenamefont
  {Goswami}}]{Kumar2010}%
  \BibitemOpen
  \bibfield  {author} {\bibinfo {author} {\bibfnamefont {D.}~\bibnamefont
  {Kumar}}, \bibinfo {author} {\bibfnamefont {C.~D.}\ \bibnamefont {Jadhav}},
  \bibinfo {author} {\bibfnamefont {P.~K.}\ \bibnamefont {Misra}},\ and\
  \bibinfo {author} {\bibfnamefont {M.}~\bibnamefont {Goswami}},\ }\bibfield
  {title} {\bibinfo {title} {{Opto-Radio Noise based True Random Number
  Generator}},\ }in\ \href {https://doi.org/10.1109/VDAT50263.2020.9190346}
  {\emph {\bibinfo {booktitle} {2020 24th International Symposium on VLSI
  Design and Test (VDAT)}}}\ (\bibinfo {year} {2020})\ pp.\ \bibinfo {pages}
  {1--5}\BibitemShut {NoStop}%
\bibitem [{\citenamefont {Prenat}\ \emph {et~al.}(2007)\citenamefont {Prenat},
  \citenamefont {El~Baraji}, \citenamefont {Guo}, \citenamefont {Sousa},
  \citenamefont {Buda-Prejbeanu}, \citenamefont {Dieny}, \citenamefont
  {Javerliac}, \citenamefont {Nozieres}, \citenamefont {Zhao},\ and\
  \citenamefont {Belhaire}}]{Prenat2007}%
  \BibitemOpen
  \bibfield  {author} {\bibinfo {author} {\bibfnamefont {G.}~\bibnamefont
  {Prenat}}, \bibinfo {author} {\bibfnamefont {M.}~\bibnamefont {El~Baraji}},
  \bibinfo {author} {\bibfnamefont {W.}~\bibnamefont {Guo}}, \bibinfo {author}
  {\bibfnamefont {R.}~\bibnamefont {Sousa}}, \bibinfo {author} {\bibfnamefont
  {L.}~\bibnamefont {Buda-Prejbeanu}}, \bibinfo {author} {\bibfnamefont
  {B.}~\bibnamefont {Dieny}}, \bibinfo {author} {\bibfnamefont
  {V.}~\bibnamefont {Javerliac}}, \bibinfo {author} {\bibfnamefont {J.-P.}\
  \bibnamefont {Nozieres}}, \bibinfo {author} {\bibfnamefont {W.}~\bibnamefont
  {Zhao}},\ and\ \bibinfo {author} {\bibfnamefont {E.}~\bibnamefont
  {Belhaire}},\ }\bibfield  {title} {\bibinfo {title} {{CMOS/magnetic hybrid
  architectures}},\ }in\ \href {https://doi.org/10.1109/ICECS.2007.4510962}
  {\emph {\bibinfo {booktitle} {{2007 14th IEEE International Conference on
  Electronics, Circuits and Systems}}}}\ (\bibinfo {organization} {IEEE},\
  \bibinfo {year} {2007})\ pp.\ \bibinfo {pages} {190--193}\BibitemShut
  {NoStop}%
\bibitem [{\citenamefont {Matsunaga}\ \emph {et~al.}(2008)\citenamefont
  {Matsunaga}, \citenamefont {Hayakawa}, \citenamefont {Ikeda}, \citenamefont
  {Miura}, \citenamefont {Hasegawa}, \citenamefont {Endoh}, \citenamefont
  {Ohno},\ and\ \citenamefont {Hanyu}}]{Matsunaga2008}%
  \BibitemOpen
  \bibfield  {author} {\bibinfo {author} {\bibfnamefont {S.}~\bibnamefont
  {Matsunaga}}, \bibinfo {author} {\bibfnamefont {J.}~\bibnamefont {Hayakawa}},
  \bibinfo {author} {\bibfnamefont {S.}~\bibnamefont {Ikeda}}, \bibinfo
  {author} {\bibfnamefont {K.}~\bibnamefont {Miura}}, \bibinfo {author}
  {\bibfnamefont {H.}~\bibnamefont {Hasegawa}}, \bibinfo {author}
  {\bibfnamefont {T.}~\bibnamefont {Endoh}}, \bibinfo {author} {\bibfnamefont
  {H.}~\bibnamefont {Ohno}},\ and\ \bibinfo {author} {\bibfnamefont
  {T.}~\bibnamefont {Hanyu}},\ }\bibfield  {title} {\bibinfo {title}
  {Fabrication of a nonvolatile full adder based on logic-in-memory
  architecture using magnetic tunnel junctions},\ }\href
  {https://doi.org/10.1143/APEX.1.091301} {\bibfield  {journal} {\bibinfo
  {journal} {Applied Physics Express}\ }\textbf {\bibinfo {volume} {1}},\
  \bibinfo {pages} {091301} (\bibinfo {year} {2008})}\BibitemShut {NoStop}%
\bibitem [{\citenamefont {Zhao}\ \emph {et~al.}(2008)\citenamefont {Zhao},
  \citenamefont {Belhaire}, \citenamefont {Chappert}, \citenamefont {Jacquet},\
  and\ \citenamefont {Mazoyer}}]{Zhao2008}%
  \BibitemOpen
  \bibfield  {author} {\bibinfo {author} {\bibfnamefont {W.}~\bibnamefont
  {Zhao}}, \bibinfo {author} {\bibfnamefont {E.}~\bibnamefont {Belhaire}},
  \bibinfo {author} {\bibfnamefont {C.}~\bibnamefont {Chappert}}, \bibinfo
  {author} {\bibfnamefont {F.}~\bibnamefont {Jacquet}},\ and\ \bibinfo {author}
  {\bibfnamefont {P.}~\bibnamefont {Mazoyer}},\ }\bibfield  {title} {\bibinfo
  {title} {{New non-volatile logic based on spin-MTJ}},\ }\href
  {https://doi.org/10.1002/pssa.200778135} {\bibfield  {journal} {\bibinfo
  {journal} {{Physica Status Solidi (a)}}\ }\textbf {\bibinfo {volume} {205}},\
  \bibinfo {pages} {1373} (\bibinfo {year} {2008})}\BibitemShut {NoStop}%
\bibitem [{\citenamefont {Kent}\ and\ \citenamefont
  {Worledge}(2015)}]{Kent2015}%
  \BibitemOpen
  \bibfield  {author} {\bibinfo {author} {\bibfnamefont {A.~D.}\ \bibnamefont
  {Kent}}\ and\ \bibinfo {author} {\bibfnamefont {D.~C.}\ \bibnamefont
  {Worledge}},\ }\bibfield  {title} {\bibinfo {title} {A new spin on magnetic
  memories},\ }\href {https://doi.org/10.1038/nnano.2015.24} {\bibfield
  {journal} {\bibinfo  {journal} {Nature Nanotechnology}\ }\textbf {\bibinfo
  {volume} {10}},\ \bibinfo {pages} {187} (\bibinfo {year} {2015})}\BibitemShut
  {NoStop}%
\bibitem [{\citenamefont {Deng}\ \emph {et~al.}(2016)\citenamefont {Deng},
  \citenamefont {Prenat}, \citenamefont {Anghel},\ and\ \citenamefont
  {Zhao}}]{Deng2016}%
  \BibitemOpen
  \bibfield  {author} {\bibinfo {author} {\bibfnamefont {E.}~\bibnamefont
  {Deng}}, \bibinfo {author} {\bibfnamefont {G.}~\bibnamefont {Prenat}},
  \bibinfo {author} {\bibfnamefont {L.}~\bibnamefont {Anghel}},\ and\ \bibinfo
  {author} {\bibfnamefont {W.}~\bibnamefont {Zhao}},\ }\bibfield  {title}
  {\bibinfo {title} {{Non-volatile magnetic decoder based on MTJs}},\ }\href
  {https://doi.org/10.1049/el.2016.2450} {\bibfield  {journal} {\bibinfo
  {journal} {Electronics Letters}\ }\textbf {\bibinfo {volume} {52}},\ \bibinfo
  {pages} {1774} (\bibinfo {year} {2016})}\BibitemShut {NoStop}%
\bibitem [{\citenamefont {Kumar}\ and\ \citenamefont
  {Thapliyal}(2019)}]{Kumar2019}%
  \BibitemOpen
  \bibfield  {author} {\bibinfo {author} {\bibfnamefont {S.~D.}\ \bibnamefont
  {Kumar}}\ and\ \bibinfo {author} {\bibfnamefont {H.}~\bibnamefont
  {Thapliyal}},\ }\bibfield  {title} {\bibinfo {title} {{Exploration of
  non-volatile MTJ/CMOS circuits for DPA-resistant embedded hardware}},\ }\href
  {https://doi.org/10.1109/TMAG.2019.2943053} {\bibfield  {journal} {\bibinfo
  {journal} {IEEE Transactions on Magnetics}\ }\textbf {\bibinfo {volume}
  {55}},\ \bibinfo {pages} {1} (\bibinfo {year} {2019})}\BibitemShut {NoStop}%
\bibitem [{\citenamefont {Barla}\ \emph
  {et~al.}(2020{\natexlab{a}})\citenamefont {Barla}, \citenamefont {Joshi},\
  and\ \citenamefont {Bhat}}]{Barla2020ALU}%
  \BibitemOpen
  \bibfield  {author} {\bibinfo {author} {\bibfnamefont {P.}~\bibnamefont
  {Barla}}, \bibinfo {author} {\bibfnamefont {V.~K.}\ \bibnamefont {Joshi}},\
  and\ \bibinfo {author} {\bibfnamefont {S.}~\bibnamefont {Bhat}},\ }\bibfield
  {title} {\bibinfo {title} {{A novel low power and reduced transistor count
  magnetic arithmetic logic unit using hybrid STT-MTJ/CMOS circuit}},\ }\href
  {https://doi.org/10.1109/ACCESS.2019.2963727} {\bibfield  {journal} {\bibinfo
   {journal} {IEEE Access}\ }\textbf {\bibinfo {volume} {8}},\ \bibinfo {pages}
  {6876} (\bibinfo {year} {2020}{\natexlab{a}})}\BibitemShut {NoStop}%
\bibitem [{\citenamefont {Barla}\ \emph
  {et~al.}(2020{\natexlab{b}})\citenamefont {Barla}, \citenamefont {Shet},
  \citenamefont {Joshi},\ and\ \citenamefont {Bhat}}]{Barla2020}%
  \BibitemOpen
  \bibfield  {author} {\bibinfo {author} {\bibfnamefont {P.}~\bibnamefont
  {Barla}}, \bibinfo {author} {\bibfnamefont {D.}~\bibnamefont {Shet}},
  \bibinfo {author} {\bibfnamefont {V.~K.}\ \bibnamefont {Joshi}},\ and\
  \bibinfo {author} {\bibfnamefont {S.}~\bibnamefont {Bhat}},\ }\bibfield
  {title} {\bibinfo {title} {{Design and Analysis of LIM Hybrid MTJ/CMOS Logic
  Gates}},\ }in\ \href {https://doi.org/10.1109/ICDCS48716.2020.243544} {\emph
  {\bibinfo {booktitle} {{2020 5th International Conference on Devices,
  Circuits and Systems (ICDCS)}}}}\ (\bibinfo {year} {2020})\ pp.\ \bibinfo
  {pages} {41--45}\BibitemShut {NoStop}%
\bibitem [{\citenamefont {Vodenicarevic}\ \emph {et~al.}(2017)\citenamefont
  {Vodenicarevic}, \citenamefont {Locatelli}, \citenamefont {Mizrahi},
  \citenamefont {Friedman}, \citenamefont {Vincent}, \citenamefont {Romera},
  \citenamefont {Fukushima}, \citenamefont {Yakushiji}, \citenamefont {Kubota},
  \citenamefont {Yuasa}, \citenamefont {Tiwari}, \citenamefont {Grollier},\
  and\ \citenamefont {Querlioz}}]{Vodenicarevic2017}%
  \BibitemOpen
  \bibfield  {author} {\bibinfo {author} {\bibfnamefont {D.}~\bibnamefont
  {Vodenicarevic}}, \bibinfo {author} {\bibfnamefont {N.}~\bibnamefont
  {Locatelli}}, \bibinfo {author} {\bibfnamefont {A.}~\bibnamefont {Mizrahi}},
  \bibinfo {author} {\bibfnamefont {J.~S.}\ \bibnamefont {Friedman}}, \bibinfo
  {author} {\bibfnamefont {A.~F.}\ \bibnamefont {Vincent}}, \bibinfo {author}
  {\bibfnamefont {M.}~\bibnamefont {Romera}}, \bibinfo {author} {\bibfnamefont
  {A.}~\bibnamefont {Fukushima}}, \bibinfo {author} {\bibfnamefont
  {K.}~\bibnamefont {Yakushiji}}, \bibinfo {author} {\bibfnamefont
  {H.}~\bibnamefont {Kubota}}, \bibinfo {author} {\bibfnamefont
  {S.}~\bibnamefont {Yuasa}}, \bibinfo {author} {\bibfnamefont
  {S.}~\bibnamefont {Tiwari}}, \bibinfo {author} {\bibfnamefont
  {J.}~\bibnamefont {Grollier}},\ and\ \bibinfo {author} {\bibfnamefont
  {D.}~\bibnamefont {Querlioz}},\ }\bibfield  {title} {\bibinfo {title}
  {{Low-Energy Truly Random Number Generation with Superparamagnetic Tunnel
  Junctions for Unconventional Computing}},\ }\href
  {https://doi.org/10.1103/PhysRevApplied.8.054045} {\bibfield  {journal}
  {\bibinfo  {journal} {Phys. Rev. Applied}\ }\textbf {\bibinfo {volume} {8}},\
  \bibinfo {pages} {054045} (\bibinfo {year} {2017})}\BibitemShut {NoStop}%
\bibitem [{\citenamefont {Borders}\ \emph {et~al.}(2019)\citenamefont
  {Borders}, \citenamefont {Pervaiz}, \citenamefont {Fukami}, \citenamefont
  {Camsari}, \citenamefont {Ohno},\ and\ \citenamefont {Datta}}]{Borders2019}%
  \BibitemOpen
  \bibfield  {author} {\bibinfo {author} {\bibfnamefont {W.~A.}\ \bibnamefont
  {Borders}}, \bibinfo {author} {\bibfnamefont {A.~Z.}\ \bibnamefont
  {Pervaiz}}, \bibinfo {author} {\bibfnamefont {S.}~\bibnamefont {Fukami}},
  \bibinfo {author} {\bibfnamefont {K.~Y.}\ \bibnamefont {Camsari}}, \bibinfo
  {author} {\bibfnamefont {H.}~\bibnamefont {Ohno}},\ and\ \bibinfo {author}
  {\bibfnamefont {S.}~\bibnamefont {Datta}},\ }\bibfield  {title} {\bibinfo
  {title} {{Integer factorization using stochastic magnetic tunnel
  junctions}},\ }\href {https://doi.org/10.1038/s41586-019-1557-9} {\bibfield
  {journal} {\bibinfo  {journal} {Nature}\ }\textbf {\bibinfo {volume} {573}},\
  \bibinfo {pages} {390} (\bibinfo {year} {2019})}\BibitemShut {NoStop}%
\bibitem [{\citenamefont {Kaiser}\ \emph {et~al.}(2019)\citenamefont {Kaiser},
  \citenamefont {Rustagi}, \citenamefont {Camsari}, \citenamefont {Sun},
  \citenamefont {Datta},\ and\ \citenamefont {Upadhyaya}}]{Kaiser2019}%
  \BibitemOpen
  \bibfield  {author} {\bibinfo {author} {\bibfnamefont {J.}~\bibnamefont
  {Kaiser}}, \bibinfo {author} {\bibfnamefont {A.}~\bibnamefont {Rustagi}},
  \bibinfo {author} {\bibfnamefont {K.~Y.}\ \bibnamefont {Camsari}}, \bibinfo
  {author} {\bibfnamefont {J.~Z.}\ \bibnamefont {Sun}}, \bibinfo {author}
  {\bibfnamefont {S.}~\bibnamefont {Datta}},\ and\ \bibinfo {author}
  {\bibfnamefont {P.}~\bibnamefont {Upadhyaya}},\ }\bibfield  {title} {\bibinfo
  {title} {{Subnanosecond Fluctuations in Low-Barrier Nanomagnets}},\ }\href
  {https://doi.org/10.1103/PhysRevApplied.12.054056} {\bibfield  {journal}
  {\bibinfo  {journal} {Phys. Rev. Applied}\ }\textbf {\bibinfo {volume}
  {12}},\ \bibinfo {pages} {054056} (\bibinfo {year} {2019})}\BibitemShut
  {NoStop}%
\bibitem [{\citenamefont {Hayakawa}\ \emph {et~al.}(2021)\citenamefont
  {Hayakawa}, \citenamefont {Kanai}, \citenamefont {Funatsu}, \citenamefont
  {Igarashi}, \citenamefont {Jinnai}, \citenamefont {Borders}, \citenamefont
  {Ohno},\ and\ \citenamefont {Fukami}}]{Hayakawa2021}%
  \BibitemOpen
  \bibfield  {author} {\bibinfo {author} {\bibfnamefont {K.}~\bibnamefont
  {Hayakawa}}, \bibinfo {author} {\bibfnamefont {S.}~\bibnamefont {Kanai}},
  \bibinfo {author} {\bibfnamefont {T.}~\bibnamefont {Funatsu}}, \bibinfo
  {author} {\bibfnamefont {J.}~\bibnamefont {Igarashi}}, \bibinfo {author}
  {\bibfnamefont {B.}~\bibnamefont {Jinnai}}, \bibinfo {author} {\bibfnamefont
  {W.~A.}\ \bibnamefont {Borders}}, \bibinfo {author} {\bibfnamefont
  {H.}~\bibnamefont {Ohno}},\ and\ \bibinfo {author} {\bibfnamefont
  {S.}~\bibnamefont {Fukami}},\ }\bibfield  {title} {\bibinfo {title}
  {{Nanosecond Random Telegraph Noise in In-Plane Magnetic Tunnel Junctions}},\
  }\href {https://doi.org/10.1103/PhysRevLett.126.117202} {\bibfield  {journal}
  {\bibinfo  {journal} {Phys. Rev. Lett.}\ }\textbf {\bibinfo {volume} {126}},\
  \bibinfo {pages} {117202} (\bibinfo {year} {2021})}\BibitemShut {NoStop}%
\bibitem [{\citenamefont {Safranski}\ \emph {et~al.}(2021)\citenamefont
  {Safranski}, \citenamefont {Kaiser}, \citenamefont {Trouilloud},
  \citenamefont {Hashemi}, \citenamefont {Hu},\ and\ \citenamefont
  {Sun}}]{Safranski2021}%
  \BibitemOpen
  \bibfield  {author} {\bibinfo {author} {\bibfnamefont {C.}~\bibnamefont
  {Safranski}}, \bibinfo {author} {\bibfnamefont {J.}~\bibnamefont {Kaiser}},
  \bibinfo {author} {\bibfnamefont {P.}~\bibnamefont {Trouilloud}}, \bibinfo
  {author} {\bibfnamefont {P.}~\bibnamefont {Hashemi}}, \bibinfo {author}
  {\bibfnamefont {G.}~\bibnamefont {Hu}},\ and\ \bibinfo {author}
  {\bibfnamefont {J.~Z.}\ \bibnamefont {Sun}},\ }\bibfield  {title} {\bibinfo
  {title} {{Demonstration of Nanosecond Operation in Stochastic Magnetic Tunnel
  Junctions}},\ }\href {https://doi.org/10.1021/acs.nanolett.0c04652}
  {\bibfield  {journal} {\bibinfo  {journal} {Nano Letters}\ }\textbf {\bibinfo
  {volume} {21}},\ \bibinfo {pages} {2040} (\bibinfo {year}
  {2021})}\BibitemShut {NoStop}%
\bibitem [{\citenamefont {Parks}\ \emph {et~al.}(2018)\citenamefont {Parks},
  \citenamefont {Bapna}, \citenamefont {Igbokwe}, \citenamefont {Almasi},
  \citenamefont {Wang},\ and\ \citenamefont {Majetich}}]{Parks2018}%
  \BibitemOpen
  \bibfield  {author} {\bibinfo {author} {\bibfnamefont {B.}~\bibnamefont
  {Parks}}, \bibinfo {author} {\bibfnamefont {M.}~\bibnamefont {Bapna}},
  \bibinfo {author} {\bibfnamefont {J.}~\bibnamefont {Igbokwe}}, \bibinfo
  {author} {\bibfnamefont {H.}~\bibnamefont {Almasi}}, \bibinfo {author}
  {\bibfnamefont {W.}~\bibnamefont {Wang}},\ and\ \bibinfo {author}
  {\bibfnamefont {S.~A.}\ \bibnamefont {Majetich}},\ }\bibfield  {title}
  {\bibinfo {title} {Superparamagnetic perpendicular magnetic tunnel junctions
  for true random number generators},\ }\href
  {https://doi.org/10.1063/1.5006422} {\bibfield  {journal} {\bibinfo
  {journal} {AIP Advances}\ }\textbf {\bibinfo {volume} {8}},\ \bibinfo {pages}
  {055903} (\bibinfo {year} {2018})}\BibitemShut {NoStop}%
\bibitem [{\citenamefont {Néel}(1949)}]{Neel1949}%
  \BibitemOpen
  \bibfield  {author} {\bibinfo {author} {\bibfnamefont {L.}~\bibnamefont
  {Néel}},\ }\bibfield  {title} {\bibinfo {title} {Théorie du traînage
  magnétique des ferromagnétiques en grains fins avec application aux terres
  cuites},\ }\href@noop {} {\bibfield  {journal} {\bibinfo  {journal} {Ann.
  Géophys.}\ }\textbf {\bibinfo {volume} {5}},\ \bibinfo {pages} {99}
  (\bibinfo {year} {1949})}\BibitemShut {NoStop}%
\bibitem [{\citenamefont {Brown}(1963)}]{Brown1963}%
  \BibitemOpen
  \bibfield  {author} {\bibinfo {author} {\bibfnamefont {W.~F.}\ \bibnamefont
  {Brown}},\ }\bibfield  {title} {\bibinfo {title} {{Thermal Fluctuations of a
  Single-Domain Particle}},\ }\href {https://doi.org/10.1103/PhysRev.130.1677}
  {\bibfield  {journal} {\bibinfo  {journal} {Phys. Rev.}\ }\textbf {\bibinfo
  {volume} {130}},\ \bibinfo {pages} {1677} (\bibinfo {year}
  {1963})}\BibitemShut {NoStop}%
\bibitem [{\citenamefont {Slonczewski}(1996)}]{SLONCZEWSKI1996L1}%
  \BibitemOpen
  \bibfield  {author} {\bibinfo {author} {\bibfnamefont {J.}~\bibnamefont
  {Slonczewski}},\ }\bibfield  {title} {\bibinfo {title} {Current-driven
  excitation of magnetic multilayers},\ }\href
  {https://doi.org/https://doi.org/10.1016/0304-8853(96)00062-5} {\bibfield
  {journal} {\bibinfo  {journal} {Journal of Magnetism and Magnetic Materials}\
  }\textbf {\bibinfo {volume} {159}},\ \bibinfo {pages} {L1} (\bibinfo {year}
  {1996})}\BibitemShut {NoStop}%
\bibitem [{\citenamefont {Berger}(1996)}]{Berger1996}%
  \BibitemOpen
  \bibfield  {author} {\bibinfo {author} {\bibfnamefont {L.}~\bibnamefont
  {Berger}},\ }\bibfield  {title} {\bibinfo {title} {Emission of spin waves by
  a magnetic multilayer traversed by a current},\ }\href
  {https://doi.org/10.1103/PhysRevB.54.9353} {\bibfield  {journal} {\bibinfo
  {journal} {Phys. Rev. B}\ }\textbf {\bibinfo {volume} {54}},\ \bibinfo
  {pages} {9353} (\bibinfo {year} {1996})}\BibitemShut {NoStop}%
\bibitem [{\citenamefont {Bertotti}\ \emph {et~al.}(2009)\citenamefont
  {Bertotti}, \citenamefont {Mayergoyz},\ and\ \citenamefont
  {Serpico}}]{BERTOTTI2009271}%
  \BibitemOpen
  \bibfield  {author} {\bibinfo {author} {\bibfnamefont {G.}~\bibnamefont
  {Bertotti}}, \bibinfo {author} {\bibfnamefont {I.~D.}\ \bibnamefont
  {Mayergoyz}},\ and\ \bibinfo {author} {\bibfnamefont {C.}~\bibnamefont
  {Serpico}},\ }\bibfield  {title} {\bibinfo {title} {{Chapter 10 - Stochastic
  Magnetization Dynamics}},\ }in\ \href
  {https://doi.org/https://doi.org/10.1016/B978-0-08-044316-4.00012-8} {\emph
  {\bibinfo {booktitle} {{Nonlinear Magnetization Dynamics in Nanosystems}}}},\
  \bibinfo {series and number} {Elsevier Series in Electromagnetism},\ \bibinfo
  {editor} {edited by\ \bibinfo {editor} {\bibfnamefont {G.}~\bibnamefont
  {Bertotti}}, \bibinfo {editor} {\bibfnamefont {I.~D.}\ \bibnamefont
  {Mayergoyz}},\ and\ \bibinfo {editor} {\bibfnamefont {C.}~\bibnamefont
  {Serpico}}}\ (\bibinfo  {publisher} {Elsevier},\ \bibinfo {address}
  {Oxford},\ \bibinfo {year} {2009})\ pp.\ \bibinfo {pages}
  {271--357}\BibitemShut {NoStop}%
\bibitem [{\citenamefont {Albert}\ \emph {et~al.}(2002)\citenamefont {Albert},
  \citenamefont {Emley}, \citenamefont {Myers}, \citenamefont {Ralph},\ and\
  \citenamefont {Buhrman}}]{Albert2002}%
  \BibitemOpen
  \bibfield  {author} {\bibinfo {author} {\bibfnamefont {F.~J.}\ \bibnamefont
  {Albert}}, \bibinfo {author} {\bibfnamefont {N.~C.}\ \bibnamefont {Emley}},
  \bibinfo {author} {\bibfnamefont {E.~B.}\ \bibnamefont {Myers}}, \bibinfo
  {author} {\bibfnamefont {D.~C.}\ \bibnamefont {Ralph}},\ and\ \bibinfo
  {author} {\bibfnamefont {R.~A.}\ \bibnamefont {Buhrman}},\ }\bibfield
  {title} {\bibinfo {title} {{Quantitative Study of Magnetization Reversal by
  Spin-Polarized Current in Magnetic Multilayer Nanopillars}},\ }\href
  {https://doi.org/10.1103/PhysRevLett.89.226802} {\bibfield  {journal}
  {\bibinfo  {journal} {Phys. Rev. Lett.}\ }\textbf {\bibinfo {volume} {89}},\
  \bibinfo {pages} {226802} (\bibinfo {year} {2002})}\BibitemShut {NoStop}%
\bibitem [{\citenamefont {Slaughter}\ \emph {et~al.}(2012)\citenamefont
  {Slaughter}, \citenamefont {Rizzo}, \citenamefont {Janesky}, \citenamefont
  {Whig}, \citenamefont {Mancoff}, \citenamefont {Houssameddine}, \citenamefont
  {Sun}, \citenamefont {Aggarwal}, \citenamefont {Nagel}, \citenamefont
  {Deshpande}, \citenamefont {Alam}, \citenamefont {Andre},\ and\ \citenamefont
  {LoPresti}}]{Slaughter2012}%
  \BibitemOpen
  \bibfield  {author} {\bibinfo {author} {\bibfnamefont {J.~M.}\ \bibnamefont
  {Slaughter}}, \bibinfo {author} {\bibfnamefont {N.~D.}\ \bibnamefont
  {Rizzo}}, \bibinfo {author} {\bibfnamefont {J.}~\bibnamefont {Janesky}},
  \bibinfo {author} {\bibfnamefont {R.}~\bibnamefont {Whig}}, \bibinfo {author}
  {\bibfnamefont {F.~B.}\ \bibnamefont {Mancoff}}, \bibinfo {author}
  {\bibfnamefont {D.}~\bibnamefont {Houssameddine}}, \bibinfo {author}
  {\bibfnamefont {J.~J.}\ \bibnamefont {Sun}}, \bibinfo {author} {\bibfnamefont
  {S.}~\bibnamefont {Aggarwal}}, \bibinfo {author} {\bibfnamefont
  {K.}~\bibnamefont {Nagel}}, \bibinfo {author} {\bibfnamefont {S.~A.}\
  \bibnamefont {Deshpande}}, \bibinfo {author} {\bibfnamefont {S.~M.}\
  \bibnamefont {Alam}}, \bibinfo {author} {\bibfnamefont {T.~W.}\ \bibnamefont
  {Andre}},\ and\ \bibinfo {author} {\bibfnamefont {P.}~\bibnamefont
  {LoPresti}},\ }\bibfield  {title} {\bibinfo {title} {{High density ST-MRAM
  technology}},\ }\href {https://doi.org/10.1109/IEDM.2012.6479128} {\bibfield
  {journal} {\bibinfo  {journal} {2012 International Electron Devices Meeting}\
  ,\ \bibinfo {pages} {29.3.1}} (\bibinfo {year} {2012})}\BibitemShut {NoStop}%
\bibitem [{\citenamefont {Thomas}\ \emph {et~al.}(2015)\citenamefont {Thomas},
  \citenamefont {Jan}, \citenamefont {Le}, \citenamefont {Lee}, \citenamefont
  {Liu}, \citenamefont {Zhu}, \citenamefont {Serrano-Guisan}, \citenamefont
  {Tong}, \citenamefont {Pi}, \citenamefont {Shen}, \citenamefont {He},
  \citenamefont {Haq}, \citenamefont {Teng}, \citenamefont {Annapragada},
  \citenamefont {Lam}, \citenamefont {Wang}, \citenamefont {Zhong},
  \citenamefont {Torng},\ and\ \citenamefont {Wang}}]{Thomas2015}%
  \BibitemOpen
  \bibfield  {author} {\bibinfo {author} {\bibfnamefont {L.}~\bibnamefont
  {Thomas}}, \bibinfo {author} {\bibfnamefont {G.}~\bibnamefont {Jan}},
  \bibinfo {author} {\bibfnamefont {S.}~\bibnamefont {Le}}, \bibinfo {author}
  {\bibfnamefont {Y.}~\bibnamefont {Lee}}, \bibinfo {author} {\bibfnamefont
  {H.}~\bibnamefont {Liu}}, \bibinfo {author} {\bibfnamefont {J.}~\bibnamefont
  {Zhu}}, \bibinfo {author} {\bibfnamefont {S.}~\bibnamefont {Serrano-Guisan}},
  \bibinfo {author} {\bibfnamefont {R.}~\bibnamefont {Tong}}, \bibinfo {author}
  {\bibfnamefont {K.}~\bibnamefont {Pi}}, \bibinfo {author} {\bibfnamefont
  {D.}~\bibnamefont {Shen}}, \bibinfo {author} {\bibfnamefont {R.}~\bibnamefont
  {He}}, \bibinfo {author} {\bibfnamefont {J.}~\bibnamefont {Haq}}, \bibinfo
  {author} {\bibfnamefont {Z.}~\bibnamefont {Teng}}, \bibinfo {author}
  {\bibfnamefont {R.}~\bibnamefont {Annapragada}}, \bibinfo {author}
  {\bibfnamefont {V.}~\bibnamefont {Lam}}, \bibinfo {author} {\bibfnamefont
  {Y.}~\bibnamefont {Wang}}, \bibinfo {author} {\bibfnamefont {T.}~\bibnamefont
  {Zhong}}, \bibinfo {author} {\bibfnamefont {T.}~\bibnamefont {Torng}},\ and\
  \bibinfo {author} {\bibfnamefont {P.}~\bibnamefont {Wang}},\ }\bibfield
  {title} {\bibinfo {title} {{Solving the paradox of the inconsistent size
  dependence of thermal stability at device and chip-level in perpendicular
  STT-MRAM}},\ }in\ \href {https://doi.org/10.1109/IEDM.2015.7409773} {\emph
  {\bibinfo {booktitle} {{2015 IEEE International Electron Devices Meeting
  (IEDM)}}}}\ (\bibinfo {year} {2015})\ pp.\ \bibinfo {pages}
  {26.4.1--26.4.4}\BibitemShut {NoStop}%
\bibitem [{\citenamefont {Jinnai}\ \emph {et~al.}(2021)\citenamefont {Jinnai},
  \citenamefont {Igarashi}, \citenamefont {Shinoda}, \citenamefont {Watanabe},
  \citenamefont {Fukami},\ and\ \citenamefont {Ohno}}]{Jinnai2021}%
  \BibitemOpen
  \bibfield  {author} {\bibinfo {author} {\bibfnamefont {B.}~\bibnamefont
  {Jinnai}}, \bibinfo {author} {\bibfnamefont {J.}~\bibnamefont {Igarashi}},
  \bibinfo {author} {\bibfnamefont {T.}~\bibnamefont {Shinoda}}, \bibinfo
  {author} {\bibfnamefont {K.}~\bibnamefont {Watanabe}}, \bibinfo {author}
  {\bibfnamefont {S.}~\bibnamefont {Fukami}},\ and\ \bibinfo {author}
  {\bibfnamefont {H.}~\bibnamefont {Ohno}},\ }\bibfield  {title} {\bibinfo
  {title} {{Fast Switching Down to 3.5 ns in sub-5-nm Magnetic Tunnel Junctions
  Achieved by Engineering Relaxation Time}},\ }in\ \href
  {https://doi.org/10.1109/IEDM19574.2021.9720509} {\emph {\bibinfo {booktitle}
  {{2021 IEEE International Electron Devices Meeting (IEDM)}}}}\ (\bibinfo
  {year} {2021})\ pp.\ \bibinfo {pages} {1--4}\BibitemShut {NoStop}%
\bibitem [{\citenamefont {Rehm}\ \emph {et~al.}(2019)\citenamefont {Rehm},
  \citenamefont {Wolf}, \citenamefont {Kardasz}, \citenamefont {Pinarbasi},\
  and\ \citenamefont {Kent}}]{Rehm2019}%
  \BibitemOpen
  \bibfield  {author} {\bibinfo {author} {\bibfnamefont {L.}~\bibnamefont
  {Rehm}}, \bibinfo {author} {\bibfnamefont {G.}~\bibnamefont {Wolf}}, \bibinfo
  {author} {\bibfnamefont {B.}~\bibnamefont {Kardasz}}, \bibinfo {author}
  {\bibfnamefont {M.}~\bibnamefont {Pinarbasi}},\ and\ \bibinfo {author}
  {\bibfnamefont {A.~D.}\ \bibnamefont {Kent}},\ }\bibfield  {title} {\bibinfo
  {title} {Sub-nanosecond spin-torque switching of perpendicular magnetic
  tunnel junction nanopillars at cryogenic temperatures},\ }\href
  {https://doi.org/10.1063/1.5128106} {\bibfield  {journal} {\bibinfo
  {journal} {Applied Physics Letters}\ }\textbf {\bibinfo {volume} {115}},\
  \bibinfo {pages} {182404} (\bibinfo {year} {2019})}\BibitemShut {NoStop}%
\bibitem [{\citenamefont {Yuasa}\ \emph {et~al.}(2013)\citenamefont {Yuasa},
  \citenamefont {Fukushima}, \citenamefont {Yakushiji}, \citenamefont {Nozaki},
  \citenamefont {Konoto}, \citenamefont {Maehara}, \citenamefont {Kubota},
  \citenamefont {Taniguchi}, \citenamefont {Arai}, \citenamefont {Imamura},
  \citenamefont {Ando}, \citenamefont {Shiota}, \citenamefont {Bonell},
  \citenamefont {Suzuki}, \citenamefont {Shimomura}, \citenamefont {Kitagawa},
  \citenamefont {Ito}, \citenamefont {Fujita}, \citenamefont {Abe},
  \citenamefont {Nomura}, \citenamefont {Noguchi},\ and\ \citenamefont
  {Yoda}}]{Yuasa2013}%
  \BibitemOpen
  \bibfield  {author} {\bibinfo {author} {\bibfnamefont {S.}~\bibnamefont
  {Yuasa}}, \bibinfo {author} {\bibfnamefont {A.}~\bibnamefont {Fukushima}},
  \bibinfo {author} {\bibfnamefont {K.}~\bibnamefont {Yakushiji}}, \bibinfo
  {author} {\bibfnamefont {T.}~\bibnamefont {Nozaki}}, \bibinfo {author}
  {\bibfnamefont {M.}~\bibnamefont {Konoto}}, \bibinfo {author} {\bibfnamefont
  {H.}~\bibnamefont {Maehara}}, \bibinfo {author} {\bibfnamefont
  {H.}~\bibnamefont {Kubota}}, \bibinfo {author} {\bibfnamefont
  {T.}~\bibnamefont {Taniguchi}}, \bibinfo {author} {\bibfnamefont
  {H.}~\bibnamefont {Arai}}, \bibinfo {author} {\bibfnamefont {H.}~\bibnamefont
  {Imamura}}, \bibinfo {author} {\bibfnamefont {K.}~\bibnamefont {Ando}},
  \bibinfo {author} {\bibfnamefont {Y.}~\bibnamefont {Shiota}}, \bibinfo
  {author} {\bibfnamefont {F.}~\bibnamefont {Bonell}}, \bibinfo {author}
  {\bibfnamefont {Y.}~\bibnamefont {Suzuki}}, \bibinfo {author} {\bibfnamefont
  {N.}~\bibnamefont {Shimomura}}, \bibinfo {author} {\bibfnamefont
  {E.}~\bibnamefont {Kitagawa}}, \bibinfo {author} {\bibfnamefont
  {J.}~\bibnamefont {Ito}}, \bibinfo {author} {\bibfnamefont {S.}~\bibnamefont
  {Fujita}}, \bibinfo {author} {\bibfnamefont {K.}~\bibnamefont {Abe}},
  \bibinfo {author} {\bibfnamefont {K.}~\bibnamefont {Nomura}}, \bibinfo
  {author} {\bibfnamefont {H.}~\bibnamefont {Noguchi}},\ and\ \bibinfo {author}
  {\bibfnamefont {H.}~\bibnamefont {Yoda}},\ }\bibfield  {title} {\bibinfo
  {title} {{Future prospects of MRAM technologies}},\ }in\ \href
  {https://doi.org/10.1109/IEDM.2013.6724549} {\emph {\bibinfo {booktitle}
  {{2013 IEEE International Electron Devices Meeting}}}}\ (\bibinfo {year}
  {2013})\ pp.\ \bibinfo {pages} {3.1.1--3.1.4}\BibitemShut {NoStop}%
\bibitem [{\citenamefont {Fukushima}\ \emph {et~al.}(2014)\citenamefont
  {Fukushima}, \citenamefont {Seki}, \citenamefont {Yakushiji}, \citenamefont
  {Kubota}, \citenamefont {Imamura}, \citenamefont {Yuasa},\ and\ \citenamefont
  {Ando}}]{Fukushima2014}%
  \BibitemOpen
  \bibfield  {author} {\bibinfo {author} {\bibfnamefont {A.}~\bibnamefont
  {Fukushima}}, \bibinfo {author} {\bibfnamefont {T.}~\bibnamefont {Seki}},
  \bibinfo {author} {\bibfnamefont {K.}~\bibnamefont {Yakushiji}}, \bibinfo
  {author} {\bibfnamefont {H.}~\bibnamefont {Kubota}}, \bibinfo {author}
  {\bibfnamefont {H.}~\bibnamefont {Imamura}}, \bibinfo {author} {\bibfnamefont
  {S.}~\bibnamefont {Yuasa}},\ and\ \bibinfo {author} {\bibfnamefont
  {K.}~\bibnamefont {Ando}},\ }\bibfield  {title} {\bibinfo {title} {{Spin
  dice: A scalable truly random number generator based on spintronics}},\
  }\href {https://doi.org/10.7567/apex.7.083001} {\bibfield  {journal}
  {\bibinfo  {journal} {Applied Physics Express}\ }\textbf {\bibinfo {volume}
  {7}},\ \bibinfo {pages} {083001} (\bibinfo {year} {2014})}\BibitemShut
  {NoStop}%
\bibitem [{\citenamefont {Choi}\ \emph {et~al.}(2014)\citenamefont {Choi},
  \citenamefont {Lv}, \citenamefont {Kim}, \citenamefont {Deshpande},
  \citenamefont {Kang}, \citenamefont {Wang},\ and\ \citenamefont
  {Kim}}]{Choi2014}%
  \BibitemOpen
  \bibfield  {author} {\bibinfo {author} {\bibfnamefont {W.~H.}\ \bibnamefont
  {Choi}}, \bibinfo {author} {\bibfnamefont {Y.}~\bibnamefont {Lv}}, \bibinfo
  {author} {\bibfnamefont {J.}~\bibnamefont {Kim}}, \bibinfo {author}
  {\bibfnamefont {A.}~\bibnamefont {Deshpande}}, \bibinfo {author}
  {\bibfnamefont {G.}~\bibnamefont {Kang}}, \bibinfo {author} {\bibfnamefont
  {J.-P.}\ \bibnamefont {Wang}},\ and\ \bibinfo {author} {\bibfnamefont
  {C.~H.}\ \bibnamefont {Kim}},\ }\bibfield  {title} {\bibinfo {title} {A
  magnetic tunnel junction based true random number generator with conditional
  perturb and real-time output probability tracking},\ }in\ \href
  {https://doi.org/10.1109/IEDM.2014.7047039} {\emph {\bibinfo {booktitle}
  {2014 IEEE International Electron Devices Meeting}}}\ (\bibinfo
  {organization} {IEEE},\ \bibinfo {year} {2014})\ pp.\ \bibinfo {pages}
  {12--5}\BibitemShut {NoStop}%
\bibitem [{\citenamefont {Matsui}(1994)}]{Matsui}%
  \BibitemOpen
  \bibfield  {author} {\bibinfo {author} {\bibfnamefont {M.}~\bibnamefont
  {Matsui}},\ }\bibfield  {title} {\bibinfo {title} {{Linear Cryptanalysis
  Method for DES Cipher}},\ }in\ \href
  {https://doi.org/10.1007/3-540-48285-7_33} {\emph {\bibinfo {booktitle}
  {{Advances in Cryptology --- EUROCRYPT '93}}}},\ \bibinfo {editor} {edited
  by\ \bibinfo {editor} {\bibfnamefont {T.}~\bibnamefont {Helleseth}}}\
  (\bibinfo  {publisher} {Springer Berlin Heidelberg},\ \bibinfo {address}
  {Berlin, Heidelberg},\ \bibinfo {year} {1994})\ pp.\ \bibinfo {pages}
  {386--397}\BibitemShut {NoStop}%
\bibitem [{\citenamefont {Bassham}\ \emph {et~al.}(2010)\citenamefont
  {Bassham}, \citenamefont {Rukhin}, \citenamefont {Soto}, \citenamefont
  {Nechvatal}, \citenamefont {Smid}, \citenamefont {Leigh}, \citenamefont
  {Levenson}, \citenamefont {Vangel}, \citenamefont {Heckert},\ and\
  \citenamefont {Banks}}]{NIST}%
  \BibitemOpen
  \bibfield  {author} {\bibinfo {author} {\bibfnamefont {L.}~\bibnamefont
  {Bassham}}, \bibinfo {author} {\bibfnamefont {A.}~\bibnamefont {Rukhin}},
  \bibinfo {author} {\bibfnamefont {J.}~\bibnamefont {Soto}}, \bibinfo {author}
  {\bibfnamefont {J.}~\bibnamefont {Nechvatal}}, \bibinfo {author}
  {\bibfnamefont {M.}~\bibnamefont {Smid}}, \bibinfo {author} {\bibfnamefont
  {S.}~\bibnamefont {Leigh}}, \bibinfo {author} {\bibfnamefont
  {M.}~\bibnamefont {Levenson}}, \bibinfo {author} {\bibfnamefont
  {M.}~\bibnamefont {Vangel}}, \bibinfo {author} {\bibfnamefont
  {N.}~\bibnamefont {Heckert}},\ and\ \bibinfo {author} {\bibfnamefont
  {D.}~\bibnamefont {Banks}},\ }\href
  {https://tsapps.nist.gov/publication/get_pdf.cfm?pub_id=906762} {\bibinfo
  {title} {{Statistical Test Suite for Random and Pseudorandom Number
  Generators for Cryptographic Applications}}} (\bibinfo {year}
  {2010})\BibitemShut {NoStop}%
\bibitem [{\citenamefont {Butler}\ \emph {et~al.}(2012)\citenamefont {Butler},
  \citenamefont {Mewes}, \citenamefont {Mewes}, \citenamefont {Visscher},
  \citenamefont {Rippard}, \citenamefont {Russek},\ and\ \citenamefont
  {Heindl}}]{Butler2012}%
  \BibitemOpen
  \bibfield  {author} {\bibinfo {author} {\bibfnamefont {W.~H.}\ \bibnamefont
  {Butler}}, \bibinfo {author} {\bibfnamefont {T.}~\bibnamefont {Mewes}},
  \bibinfo {author} {\bibfnamefont {C.~K.~A.}\ \bibnamefont {Mewes}}, \bibinfo
  {author} {\bibfnamefont {P.~B.}\ \bibnamefont {Visscher}}, \bibinfo {author}
  {\bibfnamefont {W.~H.}\ \bibnamefont {Rippard}}, \bibinfo {author}
  {\bibfnamefont {S.~E.}\ \bibnamefont {Russek}},\ and\ \bibinfo {author}
  {\bibfnamefont {R.}~\bibnamefont {Heindl}},\ }\bibfield  {title} {\bibinfo
  {title} {{Switching Distributions for Perpendicular Spin-Torque Devices
  Within the Macrospin Approximation}},\ }\href
  {https://doi.org/10.1109/TMAG.2012.2209122} {\bibfield  {journal} {\bibinfo
  {journal} {IEEE Transactions on Magnetics}\ }\textbf {\bibinfo {volume}
  {48}},\ \bibinfo {pages} {4684} (\bibinfo {year} {2012})}\BibitemShut
  {NoStop}%
\bibitem [{\citenamefont {Munira}\ \emph {et~al.}(2012)\citenamefont {Munira},
  \citenamefont {Butler},\ and\ \citenamefont {Ghosh}}]{Munira2012}%
  \BibitemOpen
  \bibfield  {author} {\bibinfo {author} {\bibfnamefont {K.}~\bibnamefont
  {Munira}}, \bibinfo {author} {\bibfnamefont {W.~H.}\ \bibnamefont {Butler}},\
  and\ \bibinfo {author} {\bibfnamefont {A.~W.}\ \bibnamefont {Ghosh}},\
  }\bibfield  {title} {\bibinfo {title} {{A Quasi-Analytical Model for
  Energy-Delay-Reliability Tradeoff Studies During Write Operations in a
  Perpendicular STT-RAM Cell}},\ }\href
  {https://doi.org/10.1109/TED.2012.2198825} {\bibfield  {journal} {\bibinfo
  {journal} {IEEE Transactions on Electron Devices}\ }\textbf {\bibinfo
  {volume} {59}},\ \bibinfo {pages} {2221} (\bibinfo {year}
  {2012})}\BibitemShut {NoStop}%
\bibitem [{\citenamefont {Liu}\ \emph {et~al.}(2014)\citenamefont {Liu},
  \citenamefont {Bedau}, \citenamefont {Sun}, \citenamefont {Mangin},
  \citenamefont {Fullerton}, \citenamefont {Katine},\ and\ \citenamefont
  {Kent}}]{Liu2014}%
  \BibitemOpen
  \bibfield  {author} {\bibinfo {author} {\bibfnamefont {H.}~\bibnamefont
  {Liu}}, \bibinfo {author} {\bibfnamefont {D.}~\bibnamefont {Bedau}}, \bibinfo
  {author} {\bibfnamefont {J.}~\bibnamefont {Sun}}, \bibinfo {author}
  {\bibfnamefont {S.}~\bibnamefont {Mangin}}, \bibinfo {author} {\bibfnamefont
  {E.}~\bibnamefont {Fullerton}}, \bibinfo {author} {\bibfnamefont
  {J.}~\bibnamefont {Katine}},\ and\ \bibinfo {author} {\bibfnamefont
  {A.}~\bibnamefont {Kent}},\ }\bibfield  {title} {\bibinfo {title} {Dynamics
  of spin torque switching in all-perpendicular spin valve nanopillars},\
  }\href {https://doi.org/https://doi.org/10.1016/j.jmmm.2014.01.061}
  {\bibfield  {journal} {\bibinfo  {journal} {Journal of Magnetism and Magnetic
  Materials}\ }\textbf {\bibinfo {volume} {358-359}},\ \bibinfo {pages} {233}
  (\bibinfo {year} {2014})}\BibitemShut {NoStop}%
\bibitem [{\citenamefont {Rehm}\ \emph {et~al.}(2021)\citenamefont {Rehm},
  \citenamefont {Wolf}, \citenamefont {Kardasz}, \citenamefont {Cogulu},
  \citenamefont {Chen}, \citenamefont {Pinarbasi},\ and\ \citenamefont
  {Kent}}]{Rehm2021}%
  \BibitemOpen
  \bibfield  {author} {\bibinfo {author} {\bibfnamefont {L.}~\bibnamefont
  {Rehm}}, \bibinfo {author} {\bibfnamefont {G.}~\bibnamefont {Wolf}}, \bibinfo
  {author} {\bibfnamefont {B.}~\bibnamefont {Kardasz}}, \bibinfo {author}
  {\bibfnamefont {E.}~\bibnamefont {Cogulu}}, \bibinfo {author} {\bibfnamefont
  {Y.}~\bibnamefont {Chen}}, \bibinfo {author} {\bibfnamefont {M.}~\bibnamefont
  {Pinarbasi}},\ and\ \bibinfo {author} {\bibfnamefont {A.~D.}\ \bibnamefont
  {Kent}},\ }\bibfield  {title} {\bibinfo {title} {{Thermal Effects in
  Spin-Torque Switching of Perpendicular Magnetic Tunnel Junctions at Cryogenic
  Temperatures}},\ }\href {https://doi.org/10.1103/PhysRevApplied.15.034088}
  {\bibfield  {journal} {\bibinfo  {journal} {Physical Review Applied}\
  }\textbf {\bibinfo {volume} {15}},\ \bibinfo {pages} {034088} (\bibinfo
  {year} {2021})}\BibitemShut {NoStop}%
\bibitem [{SMC()}]{SMCF2022}%
  \BibitemOpen
  \href@noop {} {\bibinfo {title} {{See Supplemental Material at [URL will be
  inserted by publisher] for more information about the device
  characterization, more characteristics of the SMART bitstream as well as the
  NIST test.}}}\BibitemShut {Stop}%
\bibitem [{Note1()}]{Note1}%
  \BibitemOpen
  \bibinfo {note} {It should be noted during pulses the temperature of the
  device can be significantly larger than the bath temperature due to junction
  heating~\cite {Rehm2021}.}\BibitemShut {Stop}%
\bibitem [{\citenamefont {Thas}(2010)}]{thas2010comparing}%
  \BibitemOpen
  \bibfield  {author} {\bibinfo {author} {\bibfnamefont {O.}~\bibnamefont
  {Thas}},\ }\href {https://doi.org/10.1007/978-0-387-92710-7} {\emph {\bibinfo
  {title} {{Comparing Distributions}}}},\ Vol.\ \bibinfo {volume} {233}\
  (\bibinfo  {publisher} {Springer},\ \bibinfo {year} {2010})\BibitemShut
  {NoStop}%
\bibitem [{\citenamefont {Krause}\ \emph {et~al.}(2009)\citenamefont {Krause},
  \citenamefont {Herzog}, \citenamefont {Stapelfeldt}, \citenamefont
  {Berbil-Bautista}, \citenamefont {Bode}, \citenamefont {Vedmedenko},\ and\
  \citenamefont {Wiesendanger}}]{Krause2009}%
  \BibitemOpen
  \bibfield  {author} {\bibinfo {author} {\bibfnamefont {S.}~\bibnamefont
  {Krause}}, \bibinfo {author} {\bibfnamefont {G.}~\bibnamefont {Herzog}},
  \bibinfo {author} {\bibfnamefont {T.}~\bibnamefont {Stapelfeldt}}, \bibinfo
  {author} {\bibfnamefont {L.}~\bibnamefont {Berbil-Bautista}}, \bibinfo
  {author} {\bibfnamefont {M.}~\bibnamefont {Bode}}, \bibinfo {author}
  {\bibfnamefont {E.~Y.}\ \bibnamefont {Vedmedenko}},\ and\ \bibinfo {author}
  {\bibfnamefont {R.}~\bibnamefont {Wiesendanger}},\ }\bibfield  {title}
  {\bibinfo {title} {{Magnetization Reversal of Nanoscale Islands: How Size and
  Shape Affect the Arrhenius Prefactor}},\ }\href
  {https://doi.org/10.1103/PhysRevLett.103.127202} {\bibfield  {journal}
  {\bibinfo  {journal} {Phys. Rev. Lett.}\ }\textbf {\bibinfo {volume} {103}},\
  \bibinfo {pages} {127202} (\bibinfo {year} {2009})}\BibitemShut {NoStop}%
\bibitem [{\citenamefont {Bedau}\ \emph {et~al.}(2010)\citenamefont {Bedau},
  \citenamefont {Liu}, \citenamefont {Sun}, \citenamefont {Katine},
  \citenamefont {Fullerton}, \citenamefont {Mangin},\ and\ \citenamefont
  {Kent}}]{Bedau2010}%
  \BibitemOpen
  \bibfield  {author} {\bibinfo {author} {\bibfnamefont {D.}~\bibnamefont
  {Bedau}}, \bibinfo {author} {\bibfnamefont {H.}~\bibnamefont {Liu}}, \bibinfo
  {author} {\bibfnamefont {J.~Z.}\ \bibnamefont {Sun}}, \bibinfo {author}
  {\bibfnamefont {J.~A.}\ \bibnamefont {Katine}}, \bibinfo {author}
  {\bibfnamefont {E.~E.}\ \bibnamefont {Fullerton}}, \bibinfo {author}
  {\bibfnamefont {S.}~\bibnamefont {Mangin}},\ and\ \bibinfo {author}
  {\bibfnamefont {A.~D.}\ \bibnamefont {Kent}},\ }\bibfield  {title} {\bibinfo
  {title} {{Spin-transfer pulse switching: From the dynamic to the thermally
  activated regime}},\ }\href {https://doi.org/10.1063/1.3532960} {\bibfield
  {journal} {\bibinfo  {journal} {Applied Physics Letters}\ }\textbf {\bibinfo
  {volume} {97}},\ \bibinfo {pages} {262502} (\bibinfo {year}
  {2010})}\BibitemShut {NoStop}%
\end{thebibliography}%

\end{document}